\documentclass[a4paper]{article} 

\usepackage{geometry} 
\usepackage[utf8]{inputenc}

\usepackage{fancyhdr} 
\usepackage{sectsty}
\allsectionsfont{\sffamily\mdseries\upshape} 
\usepackage[title]{appendix}

\usepackage[nottoc,notlof,notlot]{tocbibind} 
\usepackage[titles,subfigure]{tocloft} 

\usepackage{graphicx} 
\usepackage{subcaption}
\usepackage[labelfont=bf,textfont=it]{caption}

\usepackage{color} 
\usepackage{floatrow}

\usepackage{pgfplots}
\pgfplotsset{compat=1.18} 

\usepackage{amsthm}
\newtheorem{theorem}{Theorem}[section]
\newtheorem{remark}{Remark}

\usepackage{amsmath}
\usepackage{amsfonts}
\usepackage{amssymb}

\usepackage{array} 

\usepackage{tabularx}
\usepackage{booktabs} 

\usepackage{todonotes}
\usepackage{verbatim} 
\usepackage{paralist} 
\usepackage{natbib}
\usepackage{tikz}
\usetikzlibrary{arrows.meta,positioning}

\numberwithin{equation}{section}

\newcommand{\bfa}[1]{\mbox{\boldmath $ #1 $}}

\title{The impact of nonheritable variation in division rates on population growth across environments}
\author{J.A. Mackenzie, A. Hillman and M.G.M. Gomes}
         
\begin{document}
  \maketitle

\begin{abstract}
Phenotypic heterogeneity is a pervasive feature of biological populations, 
yet its impact on population growth is often interpreted through changes 
in mean individual fitness alone. In this paper, we investigate how 
nonheritable variability in division rates influences the asymptotic 
growth of a population. Using a class of linear models with phenotypic 
structure, we show that variability modifies the dominant eigenvalue 
of the system in a nonlinear manner, leading to an intrinsic trade-off: 
while variability reduces population growth under favourable conditions, 
it mitigates population decline under stress. These results provide a simple mechanistic framework for understanding 
how heterogeneity influences population-level dynamics. In particular, 
they suggest that stress-dependent amplification of mutational effects 
may arise from changes in phenotypic variability, rather than from changes 
in mean fitness alone. We illustrate this mechanism using mutation 
accumulation data in \emph{Chlamydomonas reinhardtii}, where the observed 
patterns of relative fitness under increasing stress are consistent with 
increased variability within genotypes. More broadly, our analysis highlights the importance of variability as a 
determinant of population growth, and shows that some effects commonly 
attributed to changes in mean fitness may instead reflect the nonlinear 
consequences of phenotypic heterogeneity.
\end{abstract}





\section{Introduction}

Phenotypic heterogeneity is a ubiquitous feature of biological populations. 
Even in genetically identical populations under controlled environmental conditions, 
individuals can exhibit substantial variability in division rates, growth dynamics, 
and other traits. Such variability has been widely documented across microbial systems 
and is increasingly recognised as an important determinant of population-level 
behaviour \cite{GeilerSamerotte2013}.

A large body of work has interpreted phenotypic heterogeneity in the 
context of bet-hedging strategies, in which stochastic switching between 
phenotypes enhances survival in fluctuating or unpredictable 
environments \cite{Kussell2005,Veening2008,Grimbergen2015}. 
In these frameworks, heterogeneity typically arises through phenotypic 
switching mechanisms and is analysed in environments that vary in time, 
leading to a trade-off between growth and survival. Related theoretical 
work has examined population growth in the presence of phenotypic 
diversity using branching processes or structured population models, 
where the long-term growth rate is characterised by a Lyapunov exponent or 
dominant eigenvalue \cite{Dombry2011,Vindenes2015}.

From a mathematical perspective, the impact of heterogeneity on population 
growth has been studied in a variety of frameworks, including multitype branching 
processes, age-structured models, and more general structured population approaches. 
In these models, the long-term growth rate is typically determined by a dominant 
eigenvalue, which links individual-level variability to population-level 
dynamics \cite{Rees2014}. More recent work has examined how stochastic 
variability in single-cell growth rates and division times influences population 
fitness, demonstrating that heterogeneity can have nontrivial effects even when mean 
properties are fixed \cite{Cerulus2016,Genthon2025}. However, in most cases the 
relationship between phenotypic variability and population growth is implicit, 
depending on the full structure of the model. In particular, relatively little 
is known about how changes in the distribution of division rates, independent of 
mechanisms such as phenotypic switching or environmental variability, directly 
affect the dominant growth rate.

Phenotypic heterogeneity may also arise in the absence of explicit switching mechanisms. 
A substantial body of experimental and theoretical work has shown that variability can 
emerge intrinsically from stochastic processes within cells. In particular, gene 
expression is inherently noisy, due to random fluctuations in transcription and 
translation, leading to cell-to-cell variability in protein levels even in 
genetically identical populations \cite{Raser2005,Elowitz2002}. These fluctuations 
influence key physiological processes, including metabolism and cell division, and 
can therefore generate variability in growth and division rates. Additional contributions 
arise from stochasticity in cell-cycle dynamics and the partitioning of cellular 
components at division, which introduce further nonheritable differences between 
individual cells \cite{Soltani2016,Bertaux2018}. More recently advances in single-cell measurements have revealed substantial 
variability in growth rates and division times even in constant environments, 
suggesting that phenotypic heterogeneity may arise independently of environmental 
fluctuations or switching dynamics \cite{Levy2015,GeilerSamerotte2013}. Despite 
these observations, the impact of such nonheritable variability on population 
growth remains relatively poorly understood, with much of the existing work 
focusing on mean traits rather than on the full distribution of phenotypes.

In this paper, we show that a closely related trade-off arises in a much simpler 
setting. Specifically, we consider populations in which variability in division rates is 
nonheritable and arises independently at birth, and we assume a constant environment in 
which stress acts by reducing viability. In this framework, the population dynamics are 
governed by a linear system whose long-term behaviour is determined by a dominant 
eigenvalue. This allows a direct and explicit characterisation of how phenotypic 
heterogeneity influences population growth. Our analysis shows that heterogeneity in division rates modifies population growth in 
a systematic and asymmetric way. In favourable environments, variability reduces the 
asymptotic growth rate relative to a homogeneous population with the same mean 
division rate. In contrast, under sufficiently strong stress, variability reduces 
the rate of population decline, allowing the population to persist for longer. Thus, 
heterogeneity induces an intrinsic trade-off between growth and persistence, even in 
the absence of phenotypic switching or environmental fluctuations.

To establish these results, we first analyse a two-phenotype model, which admits 
an explicit spectral characterisation. We then extend the analysis to populations 
with an arbitrary number of phenotypes, before considering a continuous formulation 
in which division rates are distributed according to a probability density. In each 
case, the dominant eigenvalue determines the long-term growth rate, while the 
corresponding eigenvector (or eigenfunction) characterises the asymptotic 
composition of the population. To illustrate the biological relevance of these results, we apply the model 
to mutation accumulation data from the green alga \emph{Chlamydomonas reinhardtii} 
\cite{kraemer2016}. These experiments show that the deleterious effects of mutations 
are amplified under stress. We demonstrate that this behaviour can be interpreted not 
only in terms of changes in mean fitness, but also as arising from changes in 
phenotypic variability within genotypes. In particular, our analysis suggests 
that increases in variability may play a key role in shaping stress-dependent 
population responses.

The two-phenotype model itself follows Gomes et al. (2019), who first 
showed that nonheritable phenotypic variation can be embedded within a 
linear population model and affects fitness estimation and coexistence. 
Their analysis, however, was largely restricted to two phenotypes and 
did not characterise how the dominant eigenvalue depends explicitly on 
the strength of environmental stress, nor how it varies systematically 
with the degree of phenotypic variability. The present paper extends 
this framework in four respects. First, we derive sharp, closed-form 
bounds on the dominant eigenvalue relative to the corresponding 
homogeneous population (Theorem 2.1), establishing the growth/decay 
asymmetry as a general property of the model rather than a numerical 
observation. Second, we prove that the dominant eigenvalue depends 
monotonically on the coefficient of variation of division rates, 
with the direction of monotonicity reversing between growth and 
decay regimes (Theorem 2.2). Third, we generalise the analysis 
from two phenotypes to an arbitrary number of discrete phenotypes 
(Theorem 3.1) and subsequently to a continuous distribution of 
division rates (Theorem 4.1), including a formal characterisation 
of the growth-rate dependence on the shape of the phenotype 
distribution via mean-preserving spreads (Theorem 4.2). Fourth, 
we apply the resulting framework to mutation accumulation data in 
Chlamydomonas reinhardtii, providing a novel interpretation of 
stress-dependent fitness effects in terms of within-genotype 
variability rather than mean fitness alone. Together, these 
extensions place the growth/persistence trade-off identified 
informally in Gomes et al. (2019) on a rigorous and considerably 
more general footing.

\section{Two-phenotype model}
We consider the two-phenotype model introduced in \cite{gomes2019effects}. 
The population consists of two phenotypes with number densities $N_1(t)$ and 
$N_2(t)$ at time $t$, and corresponding division rates $\mu_1$ and $\mu_2$. 
Without loss of generality, we assume $\mu_2 < \mu_1$. Upon division, each daughter is assigned phenotype 1 with probability $p$ and phenotype 2 
with probability $1-p$, independently of the parent cell. Therefore, phenotype assignment is nonheritable 
and does not depend on environmental conditions, but instead reflects intrinsic 
variability arising from stochastic cellular processes. As illustrated 
in Figure~\ref{fig:nonheritable_variation}, we assume that environmental stress acts at birth, 
with each daughter cell surviving with probability $1-\sigma$.

\begin{figure}
\centering

\begin{tikzpicture}[
    >=Latex,
    thick,
    node distance=2cm,
    cell/.style={
        circle, draw, fill=gray!10, minimum size=9mm
    },
    dead/.style={
        circle, draw, fill=gray!40, minimum size=9mm
    },
    edge/.style={-Latex},
    dashededge/.style={dashed,-Latex},
    lab/.style={fill=white, inner sep=1pt}
]

\node[cell] (M) at (0,0) {};
\node[above=4pt of M] {\small Mother cell};

\node[
    align=center,
    fill=white,
    inner sep=2pt
] at (4,-0.5) {\scriptsize $\sigma = \text{non-survival probability}$};

\node[cell] (D1) at (-3,-3) {};
\node[cell] (D2) at (3,-3) {};

\node[below=4pt of D1] {\small Daughter 1};
\node[below=4pt of D2] {\small Daughter 2};

\node[dead] (ND1) at (-5,-3) {};
\node[dead] (ND2) at (5,-3) {};

\draw[edge] (M) -- (D1)
    node[pos=0.5, left=-35pt, lab] {$1-\sigma$};

\draw[edge] (M) -- (D2)
    node[pos=0.5, right=-35pt, lab] {$1-\sigma$};

\draw[dashededge] (M) -- (ND1)
    node[pos=0.5, left=10pt, lab] {$\sigma$};

\draw[dashededge] (M) -- (ND2)
    node[pos=0.5, right=10pt, lab] {$\sigma$};

\node[align=center] at (-3,-5.0) {
\small
$\mathbb{P}(\text{phenotype 1}) = p$\\
$\mathbb{P}(\text{phenotype 2}) = 1-p$
};

\node[align=center] at (3,-5.0) {
\small
$\mathbb{P}(\text{phenotype 1}) = p$\\
$\mathbb{P}(\text{phenotype 2}) = 1-p$
};

\end{tikzpicture}

\caption{
Schematic illustration of nonheritable phenotypic variation under stress.
Each daughter independently survives with probability $1-\sigma$;
conditional on survival, phenotypes are assigned independently
with probabilities $p$ and $1-p$.
}
\label{fig:nonheritable_variation}
\end{figure}
Under these assumptions, the population dynamics are governed by the system
\begin{equation}
\frac{dN_{1}}{dt}=\beta p(\mu_{1}N_{1}+\mu_{2}N_{2})-\mu_{1}N_{1},
\label{eq:n1eq}
\end{equation}
\begin{equation}
\frac{dN_{2}}{dt}=\beta (1-p)(\mu_{1}N_{1}+\mu_{2}N_{2})-\mu_{2}N_{2},
\label{eq:n2eqn}
\end{equation}
where $\beta=2(1-\sigma)$ and $0 \leq \sigma \leq 1$ characterises the strength of 
the environmental stress. This system can be written in vector form as
\begin{equation}
\frac{d}{dt}
\begin{bmatrix} N_{1} \\ N_{2} \end{bmatrix}
=
\begin{bmatrix}
(\beta p-1)\mu_{1} & \beta p \mu_{2} \\
\beta(1-p)\mu_{1} & (\beta(1-p)-1)\mu_{2}
\end{bmatrix}
\begin{bmatrix} N_{1} \\ N_{2} \end{bmatrix},
\label{eq:ode2ph}
\end{equation}
or equivalently,
\begin{equation}
\dot{\bfa N} = A \bfa N,
\end{equation}
where $\bfa N(t)=[N_{1}(t),N_{2}(t)]^{T}$. The solution can be expressed in terms of the eigenvalues $\lambda^{\pm}$ of $A$ as
\[
\bfa N(t)=C_{1}e^{\lambda^{+}t}\bfa v^{+}
+ C_{2}e^{\lambda^{-}t}\bfa v^{-},
\]
where $\bfa v^{\pm}$ are the corresponding eigenvectors and the 
constants are determined by the initial conditions. The asymptotic behaviour of the system 
is governed by the eigenvalues of $A$, with the dominant eigenvalue $\lambda^{+}$ determining 
the population growth rate. A natural point of comparison is the corresponding homogeneous 
model obtained by replacing the two division rates with their mean 
\[
\bar{\mu} = p\mu_{1} + (1-p)\mu_{2},
\]
for which the population grows at rate $(\beta-1)\bar{\mu}$. The following result compares the growth rate of the heterogeneous system with this homogeneous benchmark and characterises the location of the eigenvalues.

\begin{theorem}
\label{th:2p}
Let $\lambda^\pm$ denote the eigenvalues of $A$, with $\lambda^+ \ge \lambda^-$. Then the dominant eigenvalue $\lambda^+$, which determines the population growth rate, satisfies the following properties.

\smallskip
\noindent
(i) If $1<\beta\le 2$, then the population grows exponentially and
\[
\lambda^{-} < 0 < (\beta-1)\mu_{2} < \lambda^{+} < (\beta-1)\bar{\mu}.
\]

\noindent
(ii) If $\beta=1$, then $\lambda^{+}=0$ and $\lambda^{-}<0$.

\noindent
(iii) If $0 \le \beta < 1$, then the population declines and
\[
\lambda^{-} < (\beta-1)\bar{\mu} < \lambda^{+} < (\beta-1)\mu_{2} < 0.
\]
In particular, heterogeneity in division rates reduces the population growth rate relative to the 
homogeneous system with mean division rate $\bar{\mu}$ when $1<\beta\le 2$, and 
slows population decay when $0 \leq \beta < 1$.
\end{theorem}

\begin{figure}[h!]

\begin{center}
\scalebox{0.45}{\includegraphics{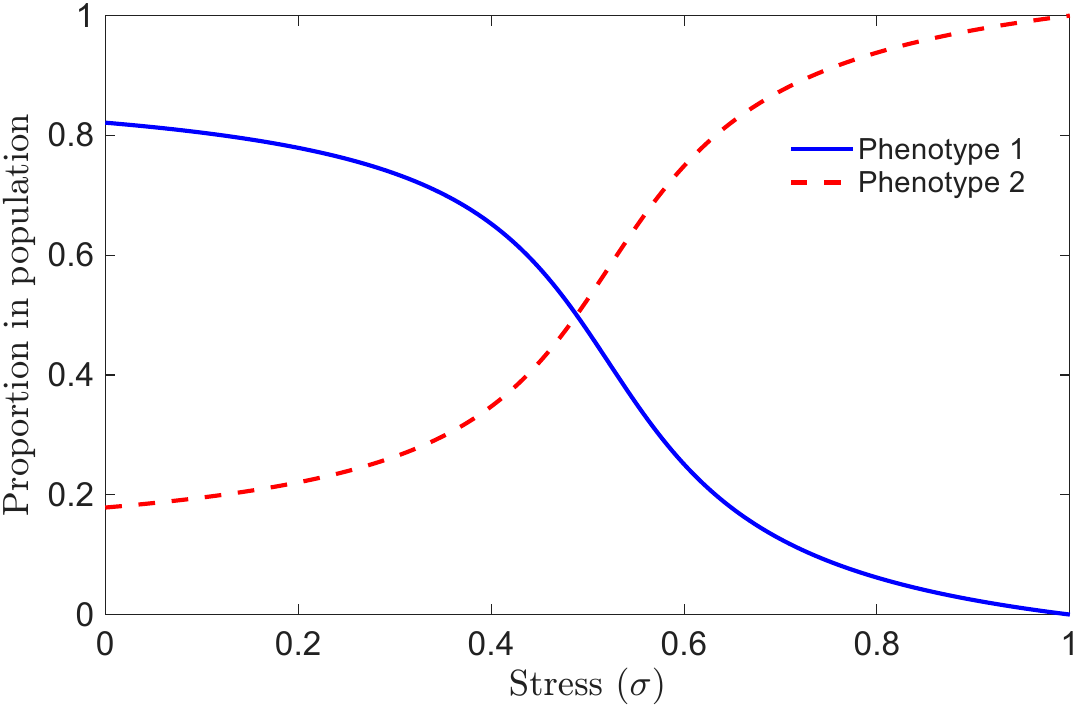}}
\end{center}
\caption{
Asymptotic phenotype proportions as a function of the stress parameter $\sigma$. Fixed parameters are $\mu_1=1$, $\mu_2=0.1$ and $p=0.9$. The proportions are obtained from the dominant eigenvector of the system matrix $A$. 
}
\label{fig:ratio_grow_kill}
\end{figure}

\begin{proof}
The eigenvalues $\lambda^{\pm}$ are the roots of the quadratic characteristic polynomial
\begin{equation}
 {\cal P}_{A}(\lambda) =  \lambda^{2}-[(\beta p-1)\mu_{1}-(1-\beta(1-p))\mu_{2}]\lambda +(1-\beta)\mu_{1}\mu_{2}.
\end{equation}
It is convenient to rewrite the linear coefficient as
\[
(\beta p-1)\mu_{1}-(1-\beta(1-p))\mu_{2}=(\beta-1)\bar{\mu}-[(1-p)\mu_{1}+p\mu_{2}]=(\beta-1)\bar{\mu}-Y,
\]
where $Y=(1-p)\mu_{1}+p\mu_{2}$. The eigenvalues therefore take the form 
\begin{equation}
    \lambda^{\pm}=\frac{(\beta-1)\bar{\mu}-Y\pm \sqrt{((\beta-1)\bar{\mu}-Y)^{2}+4(\beta-1)\mu_{1}\mu_{2}}}{2}.
    \label{eq:evalues2ph}
\end{equation}
When $\beta>1$, the constant term in the characteristic polynomial $(1-\beta)\mu_1\mu_2<0$,  
so we can deduce that $\lambda^{-}<0$ and $\lambda^{+}>0$. 
It follows that $\lambda^{+}<(\beta-1)\bar{\mu}$ if we can establish that the discriminant satisfies the inequality
\begin{equation}
((\beta-1)\bar{\mu}-Y)^{2}+4(\beta-1)\mu_{1}\mu_{2}<((\beta-1)\bar{\mu}+Y)^{2}
\label{eq:ineq}
\end{equation}
as we would then have  
\[
\lambda^{+}<\frac{(\beta-1)\bar{\mu}-Y+(\beta-1)\bar{\mu}+Y}{2}=(\beta-1)\bar{\mu}. 
\]
Since 
\[
((\beta-1)\bar{\mu}-Y)^{2}+4(\beta-1)\mu_{1}\mu_{2}=(\beta-1)^{2}\bar{\mu}^{2}-2(\beta-1)\bar{\mu}Y+Y^{2}+4(\beta-1)\mu_{1}\mu_{2}
\]
the inequality (\ref{eq:ineq}) follows if $\mu_{1}\mu_{2}<\bar{\mu} Y$. This is indeed true since 
\[
\bar{\mu}Y
= (p\mu_1+(1-p)\mu_2)((1-p)\mu_1+p\mu_2)
= \mu_1\mu_2 + p(1-p)(\mu_1-\mu_2)^2,
\]
which is strictly greater than \(\mu_1\mu_2\) whenever \(\mu_1 \ne \mu_2\). 
We have therefore established that $\lambda^{-}<0<\lambda^{+}<(\beta-1)\bar{\mu}$ 
when $1<\beta \leq 2$. To obtain a sharper lower bound for \(\lambda^+\), we evaluate 
the characteristic polynomial at \((\beta-1)\mu_2\). As illustrated in Figure \ref{fig:PA_left}, 
since \(P_A(\lambda^+)=0\) and \(P_A((\beta-1)\bar{\mu})>0\), it suffices to show that
\[
P_A((\beta-1)\mu_2) < 0,
\]
which implies \((\beta-1)\mu_2 < \lambda^+\). 

\begin{figure}[htbp]
    \centering
    \tikzset{
        >=Stealth,
        dot/.style={circle, fill=black, inner sep=1.5pt},
        lbl/.style={font=\small}
    }

    \begin{subfigure}[b]{0.48\textwidth}
        \centering
        \begin{tikzpicture}[scale=0.85] 
            \draw[->, thick] (-2.5,0) -- (5.5,0) node[right, lbl] {$z$};
            \draw[->, thick] (0,-2.5) -- (0,3.5) node[above, lbl] {$P_A(z)$};
            \node[below left, lbl] at (0,0) {$0$};
            
            \draw[domain=-2.2:4.15, smooth, variable=\z, blue, thick]
                plot ({\z}, {0.45*(\z+1.5)*(\z-3)})
                node[above right, lbl, xshift=-4pt] {$P_A(z)$};
        
            \node[dot, label={below left:$\lambda^-$}] at (-1.5,0) {};
            \node[dot, label={below right:$\lambda^+$}] at (3,0) {};
        
            \coordinate (A) at (1.2, 0);
            \coordinate (PA) at (1.2, -2.187);
            \draw[thick] (1.2, 0.1) -- (1.2, -0.1) node[above=4pt, lbl] {$(\beta-1)\mu_2$};
            \draw[dashed, gray] (A) -- (PA);
            \node[right, black, lbl, xshift=8pt] at (PA) {$P_A((\beta-1)\mu_2) < 0$};
        
            \coordinate (B) at (3.85, 0);
            \coordinate (PB) at (3.85, 2.046); 
            \draw[thick] (3.85, 0.1) -- (3.85, -0.1);
            \node[above right, lbl, xshift=2pt, yshift=2pt] at (3.85, 0) {$(\beta-1)\bar{\mu}$};
            
            \draw[dashed, gray] (B) -- (PB);
            \node[left, black, lbl, xshift=-2pt] at (PB) {$P_A((\beta-1)\bar{\mu}) > 0$};
        \end{tikzpicture}
        \caption{$1 < \beta \leq 2$}
        \label{fig:PA_left}
    \end{subfigure}
    \hfill 
    \begin{subfigure}[b]{0.48\textwidth}
        \centering
        \begin{tikzpicture}[scale=0.85]
            \draw[->, thick] (-5.5,0) -- (1.5,0) node[right, lbl] {$z$};
            \draw[->, thick] (0,-2.0) -- (0,3.5) node[above, lbl] {$P_A(z)$};
            \node[below right, lbl] at (0,0) {$0$};
        
            \draw[domain=-5.0:-0.3, smooth, variable=\z, blue, thick] 
                plot ({\z}, {0.5*(\z+4.5)*(\z+1.5)}) 
                node[above right, lbl, xshift=-24pt] {$P_A(z)$};
        
            \node[dot, label={above right, lbl, xshift=-4pt, yshift=2pt:$\lambda^-$}] at (-4.5,0) {};
            \node[dot, label={above left, lbl, xshift=6pt, yshift=2pt:$\lambda^+$}] at (-1.5,0) {};
        
            \coordinate (C) at (-3.2, 0);
            \coordinate (PC) at (-3.2, -1.105); 
            \draw[thick] (-3.2, 0.1) -- (-3.2, -0.1);
            \node[above right, lbl, xshift=-16pt, yshift=2pt] at (-3.2, 0) {$(\beta-1)\bar{\mu}$};
            
            \draw[dashed, gray] (C) -- (PC);
            \node[left, darkgray, lbl, xshift=-20pt] at (PC) {$P_A((\beta-1)\bar{\mu}) < 0$};
        
            \coordinate (D) at (-0.8, 0);
            \coordinate (PD) at (-0.8, 1.295);
            \draw[thick] (-0.8, 0.1) -- (-0.8, -0.1) node[below=8pt, lbl, xshift=-4pt] {$(\beta-1)\mu_2$};
            
            \draw[dashed, gray] (D) -- (PD);
            \node[right, black, lbl, xshift=-90pt] at (PD) {$P_A((\beta-1)\mu_2) > 0$};
        \end{tikzpicture}
        \caption{$0 < \beta \leq 1$}
        \label{fig:PA_right}
    \end{subfigure}

    \caption{Schematic of the characteristic polynomial $P_A(z)$ under different regimes of $\beta$, showing the eigenvalues $\lambda^{-}$ and $\lambda^{+}$, along with the test points $(\beta-1)\mu_2$ and $(\beta-1)\bar{\mu}$ which bound $\lambda^{+}$.}
    \label{fig:PA_combined_schematic}
\end{figure}
\noindent
A direct calculation shows that
\begin{equation}
{\cal P}_{A}((\beta-1)\mu_{2})=(\beta-1)\mu_{2}(\beta p(\mu_{2}-\mu_{1})).
\label{eq:PA_mu2}
\end{equation}
It follows that ${\cal P}_{A}((\beta-1)\mu_{2})<0$ when $1<\beta \leq 2$, 
which establishes that $\lambda^{+}> (\beta-1)\mu_{2}$. When $\beta=1$ it's clear 
from (\ref{eq:evalues2ph}) that $\lambda^{-}=-2Y<0$ and $\lambda^{+}=0$. When $\beta<1$ the discriminant  
\begin{eqnarray}
((\beta-1)\bar{\mu}-Y)^{2}+4(\beta-1)\mu_{1}\mu_{2}& = & (\beta-1)^{2}\bar{\mu}^{2}-2(\beta-1)\bar{\mu}Y+Y^{2}+4(\beta-1)\mu_{1}\mu_{2} \nonumber \\
& > & ((\beta-1)\bar{\mu}+Y)^{2} 
\end{eqnarray}
as $\mu_{1}\mu_{2}<\bar{\mu}Y$. We therefore have $\lambda^{-}<-2Y<0$ and 
\[
0>\lambda^{+}> \frac{(\beta-1)\bar{\mu}-Y+(\beta-1)\bar{\mu}+Y}{2}=(\beta-1)\bar{\mu}.
\]
We can obtain a sharper upper bound for $\lambda^{+}$ by evaluating the characteristic 
polynomial at $(\beta-1)\mu_{2}$. 
Since ${\cal P}_{A}(\lambda^{+})=0$ and ${\cal P}_{A}((\beta-1)\bar{\mu})<0$, it is sufficient to establish that 
${\cal P}_{A}((\beta-1)\mu_{2})>0$ as illustrated in Figure \ref{fig:PA_right}.
From (\ref{eq:PA_mu2}) is follows that ${\cal P}_{A}((\beta-1)\mu_{2})>0$ when $0\leq \beta <1$ and this completes the proof.
\end{proof}
\begin{remark}
The above theorem focuses on the comparison of $\lambda^{+}$ with the growth rate of a homogeneous model with the division rate $\bar{\mu}$. In the stress-free situation $\beta=2$ with $p=1/2$ it is clear that $\lambda^{+}=\sqrt{\mu_{1}\mu_{2}}$. The asymptotic growth rate is therefore equal to the geometric average, $M_{g}$, of $\mu_{1}$ and $\mu_{2}$. It is well known that $M_{g}<M_{a}$, where $M_{a}$ is the arithmetic average between two positive quantities. So, when there is an equal probability that any newly born cell can belong to either phenotype, we observe the counter-intuitive behaviour that the heterogeneous population growth is less than that expected based simply on the arithmetic average of division rates.  
\end{remark}
\begin{remark}
In \cite{hashimoto2016noise} the authors describe a growth rate 
gain of a heterogeneous model of a growing population of bacteria compared to an equivalent homogeneous model. Their model is based on a heterogeneous distribution of cellular subdivision times rather than a distribution of division rates. To relate their model to that considered here we can define times to subdivision as the reciprocal of division rates. We can then consider a homogeneous model with a division rate which is given by the inverse of the arithmetic average of the subdivision times. That is we can compare with the growth rate obtained using the harmonic average 
\[
M_{h}=\frac{2\mu_{1}\mu_{2}}{\mu_{1}+\mu_{2}},
\]
where 
\[
\frac{1}{M_{h}}=\frac{1}{2}\left ( \frac{1}{\mu_{1}}+\frac{1}{\mu_{2}} \right ).
\]
It is well known that $M_{h}<M_{g}$, so if one compares the population growth rate of a heterogeneous model with an equivalent homogeneous model based on the inverse of an arithmetic average of subdivision times, then one would come to conclusion that there is an increase in the growth rate of the heterogeneous model. To avoid confusion, it is therefore crucial to correctly define an appropriate mean division rate in order to compare the growth rates of heterogeneous and homogeneous models.  
\end{remark}
While Theorem~\ref{th:2p} characterises the effect of heterogeneity on the overall 
population growth rate, additional insight can be obtained by examining the asymptotic 
composition of the population. This composition is determined by the dominant eigenvector 
of $A$, corresponding to the eigenvalue $\lambda^{+}$. Since $A$ is a Metzler 
matrix (i.e. has nonnegative off-diagonal entries), its dominant eigenvector can be 
chosen to have strictly positive components (see, e.g., \cite{smith1995monotone}), and 
hence defines well-posed phenotype proportions. 
Figure~\ref{fig:ratio_grow_kill} shows these asymptotic 
proportions as a function of the stress parameter $\sigma$ for 
fixed values of $p$ and division rates $\mu_{1}$ and $\mu_{2}$. At low stress levels, 
the population is dominated by the faster-dividing phenotype, consistent with the 
higher growth rate identified in Theorem~\ref{th:2p}. As the stress level increases, 
however, the proportion of slower-dividing cells increases, eventually leading to 
their dominance in high-stress regimes. This reflects a selection effect induced by 
environmental stress: conditions that reduce population growth also 
favour phenotypes with lower division rates, thereby reshaping the asymptotic 
population structure.

We note that this selection effect, whereby environmental stress leads to a shift 
towards slower-dividing phenotypes, is not specific to the modelling choice used here. 
In Appendix~A, we consider an alternative formulation in which stress acts during the 
lifetime of each cell rather than at birth. Numerical results for this model exhibit 
the same qualitative behaviour, indicating that the dependence of the population dynamics on 
heterogeneity is robust to the mechanism by which stress is incorporated.

\subsection{Dependency of growth rate on the coefficient of variation}
\label{sec:cv}
We now examine how the population growth rate depends on the level of heterogeneity 
in division rates. In particular, we study how the dominant eigenvalue $\lambda^{+}$ 
varies with the coefficient of variation (CV), while keeping the mean division rate fixed. 
The coefficient of variation in the division rates
\begin{equation}
{\rm CV}=\frac{\sqrt{p(\mu_{1}-\bar{\mu})^2+(1-p)(\mu_{2}-\bar{\mu})^{2}}}{\bar{\mu}}.
\label{eq:defncv}
\end{equation}
To isolate the effect of heterogeneity, we fix the mean division rate $\bar{\mu}$ and consider variations in $\mu_1$ and $\mu_2$ that 
change the coefficient of variation while preserving $\bar{\mu}$. To investigate this dependence, we parameterise the division rates in terms of the mean $\bar{\mu}$, the probability $p$, and the coefficient of variation. A convenient representation is
\begin{equation}
\mu_{1}=\bar{\mu}\left(1+\phi\,\mathrm{CV}\right), \quad
\mu_{2}=\bar{\mu}\left(1-\frac{\mathrm{CV}}{\phi}\right),
\label{eq:detmu}
\end{equation}
where $\phi=\sqrt{(1-p)/p}$ and to ensure $\mu_{2}>0$ we require ${\rm CV}< \phi$. This 
choice ensures that $\bar{\mu}$ remains fixed while the 
variability in division rates is controlled by CV.
 
\begin{figure}[h!]
\begin{center}
 \leavevmode
 \mbox{
\begin{minipage}{0.33\textwidth}
\begin{center}
\scalebox{0.3}{\includegraphics{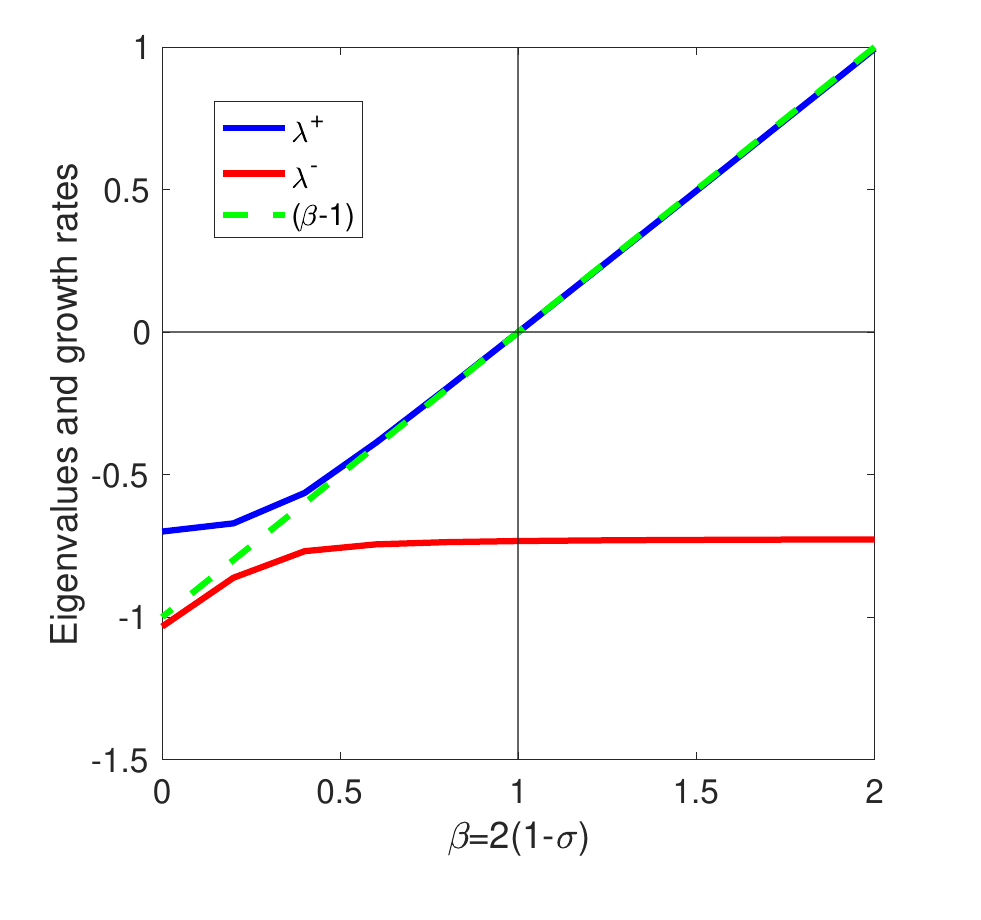}}\\
 {\small CV=0.1}
\end{center}
 \end{minipage}
\begin{minipage}{0.33\textwidth}
\begin{center}
\scalebox{0.3}{\includegraphics{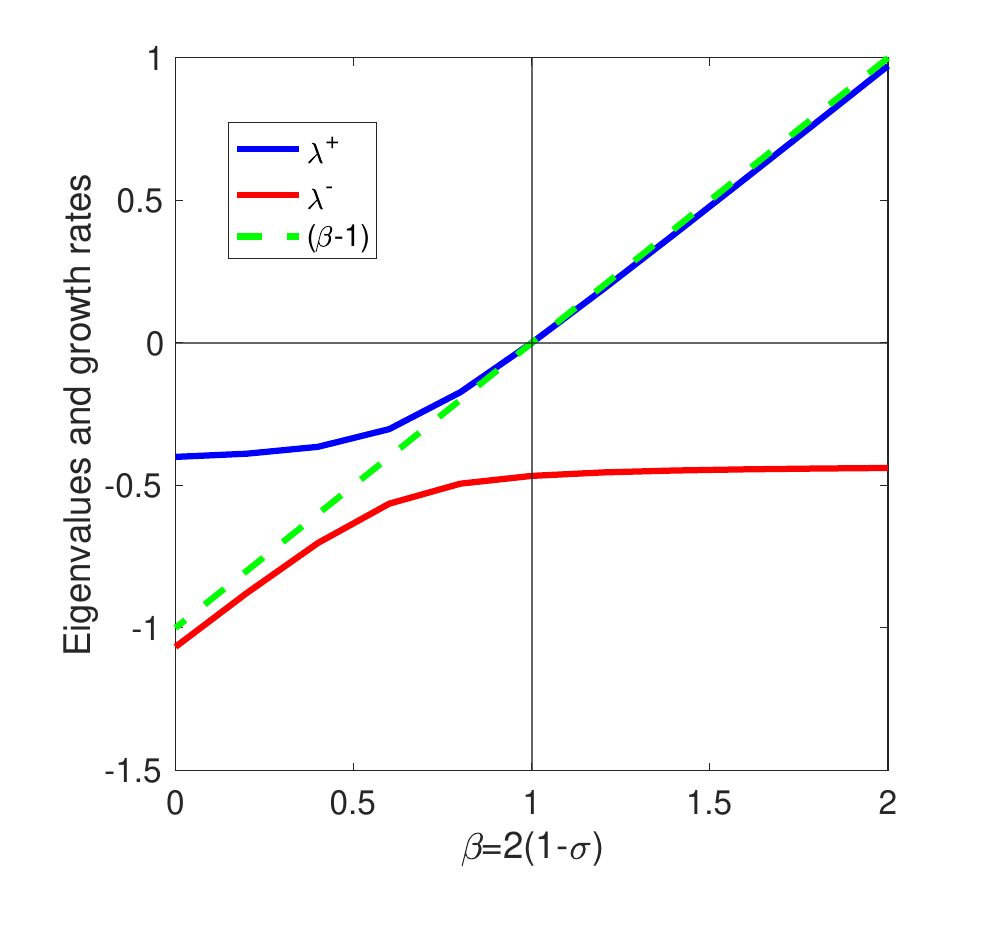}}\\
 {\small CV=0.2}
\end{center}
 \end{minipage}
\begin{minipage}{0.33\textwidth}
\begin{center}
\scalebox{0.3}{\includegraphics{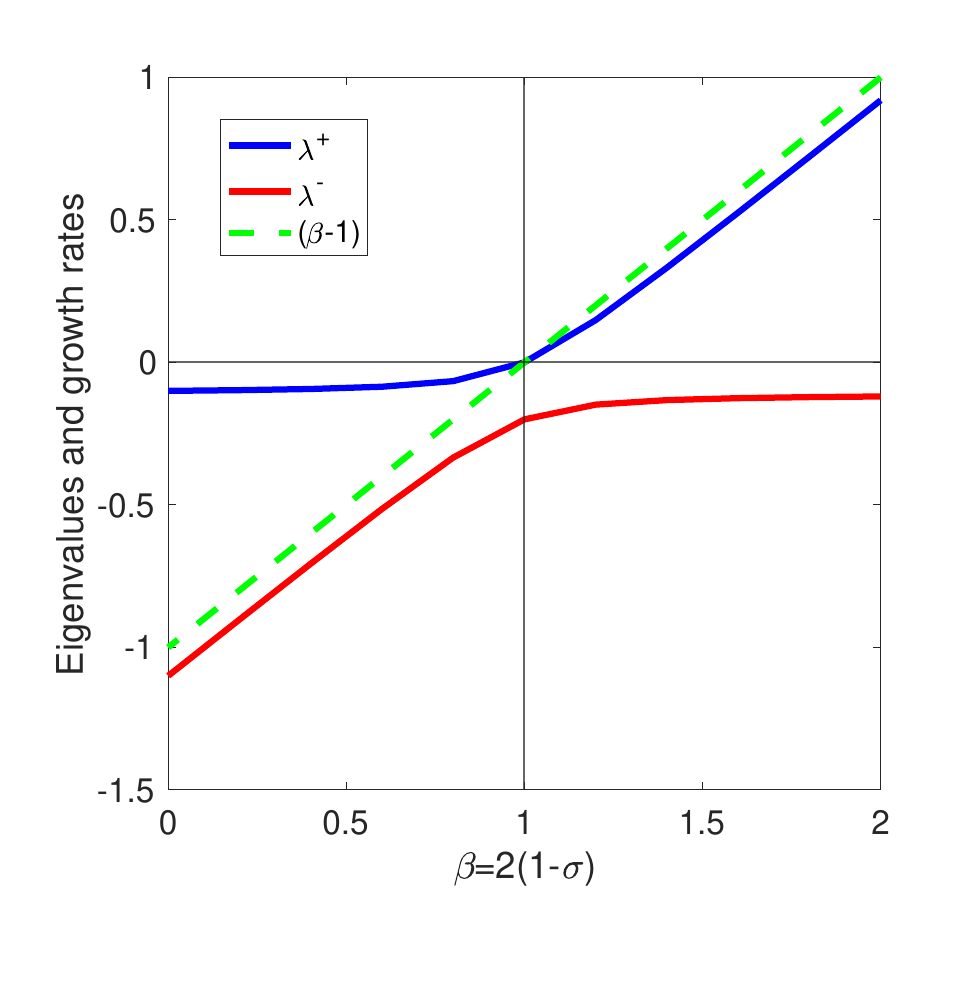}}\\
 {\small CV=0.3}
\end{center}
 \end{minipage} 
}
\end{center}
\caption{Eigenvalues $\lambda^{\pm}$ of the two-phenotype model as a function of the stress parameter $\beta=2(1-\sigma)$, shown for three levels of heterogeneity (coefficient of variation $\mathrm{CV} = 0.1, 0.2, 0.3$). 
The dashed line shows the growth rate of the corresponding homogeneous population with mean division rate $\bar{\mu}$. The dominant eigenvalue $\lambda^{+}$, which determines the asymptotic population growth rate, lies below the homogeneous growth rate for $\beta > 1$ (growth regime) and above it for $\beta < 1$ (decay regime), with the discrepancy increasing as variability (CV) increases.
}
\label{fig:stress_2p}
\end{figure}

Figure~\ref{fig:stress_2p} illustrates how the eigenvalues $\lambda^\pm$ vary with the 
stress parameter $\beta$ for different levels of heterogeneity. Throughout, we take $p=0.9$ and 
$\bar{\mu}=1$, with the division rates $\mu_1$ and $\mu_2$ determined from the 
parameterisation \eqref{eq:detmu}. Consistent with Theorem~\ref{th:2p}, the dominant eigenvalue $\lambda^{+}$ lies 
below the homogeneous growth rate $(\beta-1)\bar{\mu}$ in the growth regime ($\beta>1$) 
and above it in the decay regime ($\beta<1$). As the coefficient of variation increases, the discrepancy between $\lambda^{+}$ 
and $(\beta-1)\bar{\mu}$ becomes more pronounced. In particular, higher variability leads 
to reduced population growth under favourable conditions and a slower rate of decline under 
strong stress. This behaviour is most evident at large CV and high stress levels, where the 
growth rate approaches that of the slower-dividing phenotype.

\begin{figure}[h!]
\begin{center}
 \leavevmode
 \mbox{
\begin{minipage}{0.5\textwidth}
\begin{center}
\scalebox{0.45}{\includegraphics{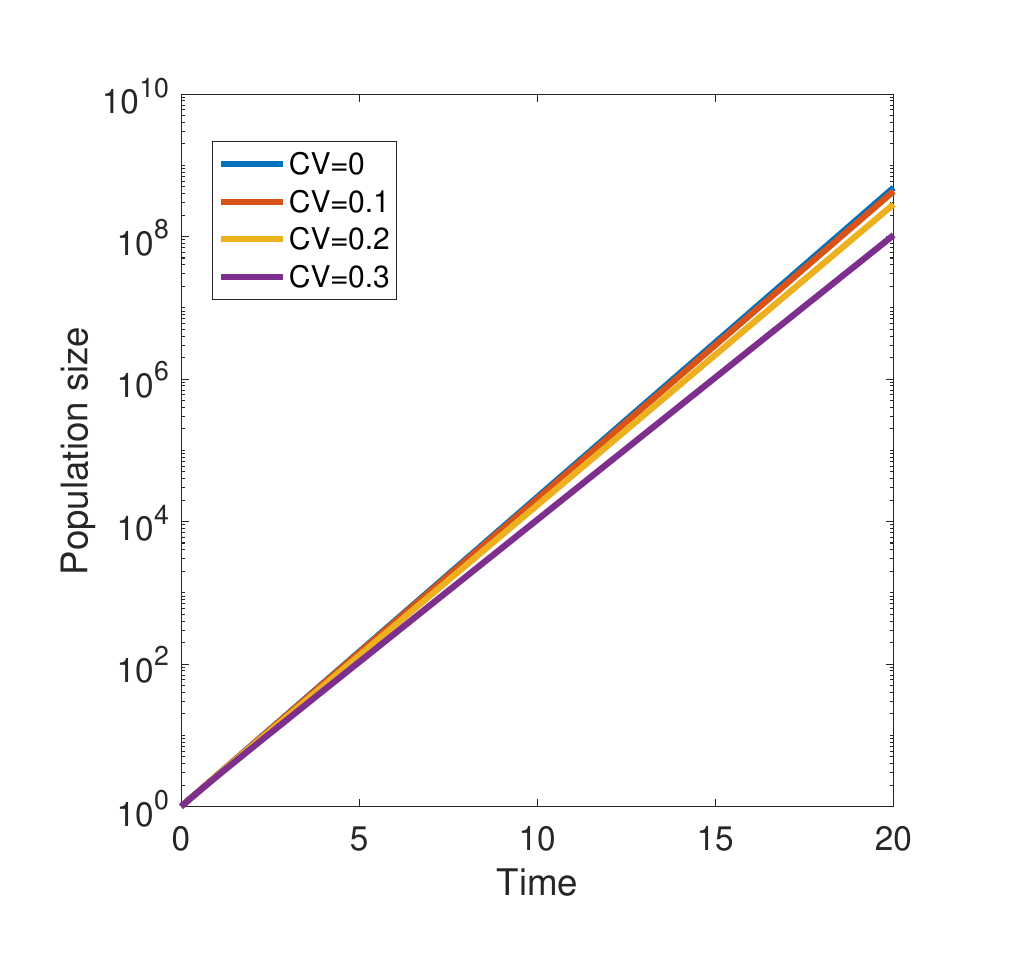}}\\
 {\small (a) Zero stress  ($\beta=2$).}
\end{center}
 \end{minipage}
\begin{minipage}{0.5\textwidth}
\begin{center}
\scalebox{0.45}{\includegraphics{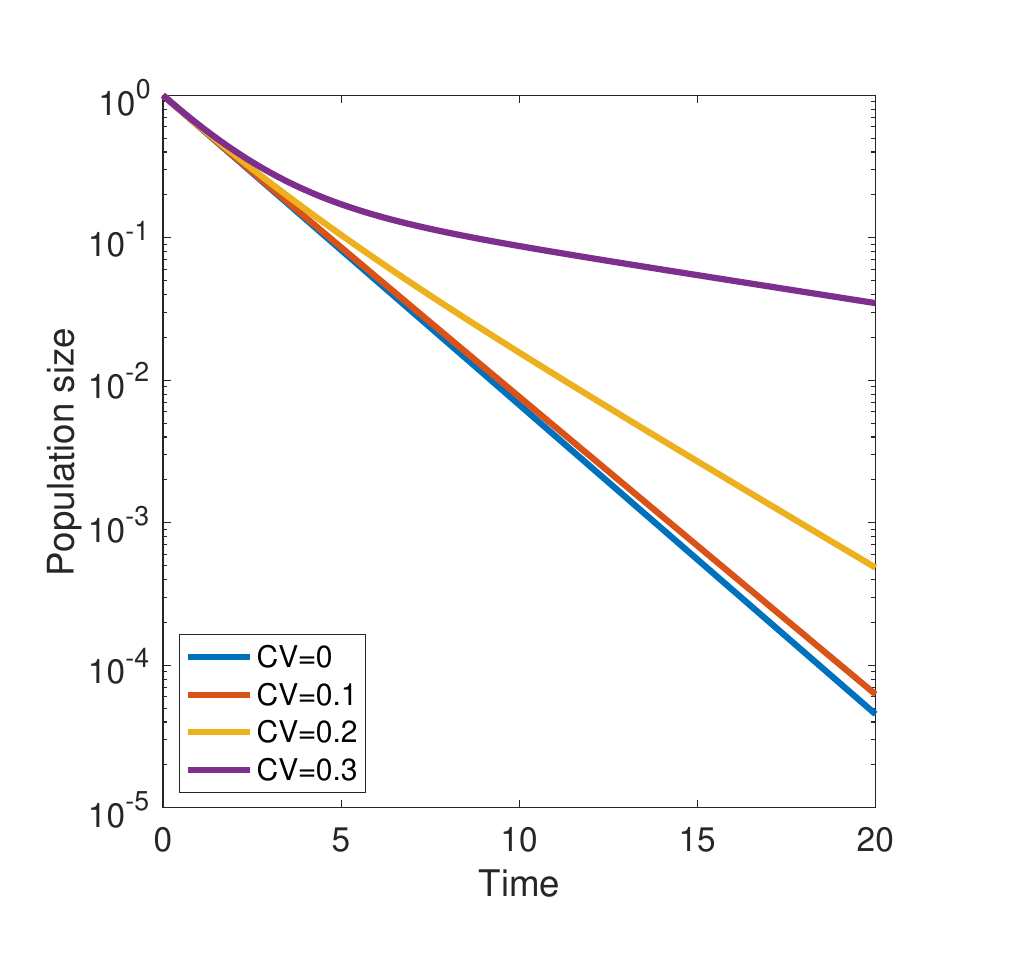}}\\
 {\small (b) High stress ($\beta=0.25$).}
\end{center}
 \end{minipage}
}
\end{center}
\caption{
Time evolution of population size for the homogeneous model (CV = 0) and the 
heterogeneous two-phenotype model for increasing levels of variability 
(CV = 0.1, 0.2, 0.3), with identical mean division rate. Panels show (a) 
low-stress conditions ($\beta=2$) and (b) high-stress conditions ($\beta=0.25$). 
In favourable environments (a), all populations initially grow at similar rates 
but diverge asymptotically, with higher variability leading to slower growth. 
Under strong stress (b), populations decay, but increased variability leads to slower 
long-term decline due to selection for low-division-rate phenotypes.}
\label{fig:2pgrow_kill}
\end{figure}

Figure~\ref{fig:2pgrow_kill} illustrates the time evolution of the homogeneous model (CV = 0) and the total population $N_{1}(t)+N_{2}(t)$ in the two-phenotype model, 
for $p=0.9$, $\bar{\mu}=1$, and three non-zero values of CV. Under low-stress 
conditions ($\beta=2$), all populations initially grow at identical rates, but 
diverge asymptotically, with higher variability leading to slower growth. Under 
strong stress ($\beta=0.25$), the populations decay, with higher variability leading 
to a slower long-term rate of decline. This behaviour reflects a stress-induced selection 
effect: increasing stress suppresses faster-dividing cells, leading to a population 
dominated asymptotically by slower-dividing phenotypes and hence governed by slower 
growth or decay rates. These observations suggest a monotonic dependence of the dominant eigenvalue $\lambda^{+}$ on the level of heterogeneity. In particular, increasing variability reduces $\lambda^{+}$ in the growth regime and increases it (i.e. makes it less negative) in the decay regime. The following result establishes this dependence 
rigorously in terms of the coefficient of variation.

\begin{theorem}
\label{th:2pcv}
Fix the mean division rate $\bar{\mu}$ and the probability $p$, and let $\lambda^{+}$ denote the dominant eigenvalue of the two-phenotype system. Then the dependence of $\lambda^{+}$ on the coefficient of variation satisfies
\[
\frac{\partial \lambda^{+}}{\partial\,\mathrm{CV}} < 0,
\qquad 1<\beta\le 2,
\]
and
\[
\frac{\partial \lambda^{+}}{\partial\,\mathrm{CV}} > 0,
\qquad 0 \le \beta < 1.
\]
In particular, increasing heterogeneity reduces the population growth rate in the growth regime and reduces the rate of population decay in the decay regime.
\end{theorem}
\begin{proof}
We fix $\bar{\mu}$ and $p$, and regard $\lambda^{+}$ as a function of the coefficient of variation. From the parameterisation
\[
\mu_{2}=\bar{\mu}\left(1-\frac{\mathrm{CV}}{\phi}\right),
\qquad
\phi=\sqrt{\frac{1-p}{p}},
\]
it follows that
\[
\frac{\partial \lambda^{+}}{\partial\,\mathrm{CV}}
=
\frac{\partial \lambda^{+}}{\partial \mu_{2}}
\frac{\partial \mu_{2}}{\partial\,\mathrm{CV}}
=
-\frac{\bar{\mu}}{\phi}\frac{\partial \lambda^{+}}{\partial \mu_{2}}.
\]
Hence
\[
\operatorname{sign}\!\left(\frac{\partial \lambda^{+}}{\partial\,\mathrm{CV}}\right)
=
-\operatorname{sign}\!\left(\frac{\partial \lambda^{+}}{\partial \mu_{2}}\right).
\]
We now express the characteristic polynomial in the form
\[
\mathcal{P}_A(\lambda)=\lambda^{2}+b\lambda+c,
\]
where
\[
b=b(p,\bar{\mu},\mu_{2}),
\qquad
c=c(p,\bar{\mu},\mu_{2}).
\]
Since $\lambda^{+}$ is a root of $\mathcal{P}_A$, differentiation of the identity
\[
(\lambda-\lambda^{+})(\lambda-\lambda^{-})
=
\lambda^{2}+b\lambda+c
\]
with respect to $\mu_{2}$ gives
\[
\frac{\partial \lambda^{+}}{\partial \mu_{2}}
=
-\frac{1}{\lambda^{+}-\lambda^{-}}
\left(
\lambda^{+}\frac{\partial b}{\partial \mu_{2}}
+
\frac{\partial c}{\partial \mu_{2}}
\right).
\]
Since $\lambda^{+}>\lambda^{-}$, it follows that
\[
\operatorname{sign}\!\left(\frac{\partial \lambda^{+}}{\partial\,\mathrm{CV}}\right)
=
\operatorname{sign}\!\left(
\lambda^{+}\frac{\partial b}{\partial \mu_{2}}
+
\frac{\partial c}{\partial \mu_{2}}
\right).
\]
A direct calculation yields
\[
\lambda^{+}\frac{\partial b}{\partial \mu_{2}}
+
\frac{\partial c}{\partial \mu_{2}}
=
\frac{1}{p}
\left[
2(1-p)\bigl((\beta-1)\mu_{2}-\lambda^{+}\bigr)
+
\bigl(\lambda^{+}-(\beta-1)\bar{\mu}\bigr)
\right].
\]
We now use the bounds established in Theorem~\ref{th:2p}. If $1<\beta\le 2$, then
\[
(\beta-1)\mu_{2}<\lambda^{+}<(\beta-1)\bar{\mu},
\]
so both terms inside the brackets are negative. Hence
\[
\lambda^{+}\frac{\partial b}{\partial \mu_{2}}
+
\frac{\partial c}{\partial \mu_{2}}<0,
\]
and therefore
\[
\frac{\partial \lambda^{+}}{\partial\,\mathrm{CV}}<0.
\]
If $0\le \beta<1$, then
\[
(\beta-1)\bar{\mu}<\lambda^{+}<(\beta-1)\mu_{2},
\]
so both terms inside the brackets are positive. Hence
\[
\lambda^{+}\frac{\partial b}{\partial \mu_{2}}
+
\frac{\partial c}{\partial \mu_{2}}>0,
\]
and therefore
\[
\frac{\partial \lambda^{+}}{\partial\,\mathrm{CV}}>0.
\]
This completes the proof.
\end{proof}
Theorem~\ref{th:2pcv} shows that the dependence of the population growth rate on heterogeneity is both systematic and robust in the two-phenotype setting, 
with increasing variability reducing growth in favourable environments and reducing 
the rate of decay under strong stress. This behaviour reflects the selection mechanism 
identified above, whereby environmental conditions shift the population towards 
slower-dividing phenotypes. 

\section{$n$-phenotype model}

Motivated by the results obtained for the two-phenotype system, we now consider 
a population consisting of $n$ phenotypes with 
number densities $N_i(t)$, $i=1,\ldots,n$, and corresponding division rates $\mu_i>0$. Without loss of generality, we assume that the division rates are ordered such that
\[
0 < \mu_n < \mu_{n-1} < \cdots < \mu_2 < \mu_1.
\]
Upon division, each daughter cell is assigned phenotype $i$ with probability $p_i$, 
independently of the parent, where $p_i > 0$ and $\sum_{i=1}^n p_i = 1$. As in Section~2, 
environmental stress is incorporated through a reduction in viability at birth. Under these assumptions, the population dynamics are governed by
\[
\frac{dN_i}{dt}
=
\beta p_i \sum_{j=1}^n \mu_j N_j
-
\mu_i N_i,
\qquad i=1,\ldots,n,
\]
where $\beta = 2(1-\sigma)$ as before. This system can be written in vector form as
\[
\dot{\bfa N} = A \bfa N,
\]
where $\bfa N(t) = (N_1(t),\ldots,N_n(t))^T$ and the matrix $A$ is given by
\[
A_{ij} =
\begin{cases}
\beta p_i \mu_j, & i \ne j, \\
(\beta p_i -1)\mu_i, & i=j.
\end{cases}
\]
As in the two-phenotype case, the asymptotic behaviour of the population is determined by the 
eigenvalues of $A$, with the dominant eigenvalue governing the long-term growth rate. The 
following result generalises Theorem~\ref{th:2p} to the $n$-phenotype setting.

\begin{theorem}
\label{th:3p}
Let $\lambda_1 < \lambda_2 < \cdots < \lambda_n$ denote the eigenvalues of $A$, with $\lambda_n$ the dominant eigenvalue. Then
\[
-\mu_1 < \lambda_1 < -\mu_2 < \lambda_2 < \cdots < -\mu_{n-1} < \lambda_{n-1} < -\mu_n.
\]
Moreover, the dominant eigenvalue satisfies
\[
0 < (\beta-1)\mu_n < \lambda_n < (\beta-1)\bar{\mu},
\qquad 1<\beta\le 2,
\]
\[
\lambda_n = 0,
\qquad \beta = 1,
\]
and
\[
{\rm max}[-\mu_{n},(\beta-1)\bar{\mu}] < \lambda_n < (\beta-1)\mu_n < 0,
\qquad 0<\beta<1,
\]
where
\[
\bar{\mu}=\sum_{i=1}^n p_i\mu_i.
\]
\end{theorem}

\begin{proof}
We first show that $\lambda \neq -\mu_i$ for all $i = 1, \dots, n$ by contradiction.
Suppose that $\lambda = -\mu_k$ for some index $k$. Evaluating the characteristic 
polynomial directly at this point eliminates all but one term in the expansion, yielding:
\[
P_A(-\mu_k) = (-1)^{n+1} \beta p_k \mu_k \prod_{j \neq k} (\mu_j - \mu_k).
\]
Since $\beta > 0$, $p_k > 0$, $\mu_k > 0$, and the 
elements $\mu_i$ are strictly distinct ($0 < \mu_n < \dots < \mu_1$), it follows 
that $P_A(-\mu_k) \neq 0$ and hence $-\mu_{k}$ is not an eigenvalue of $A$. 

The matrix $A$ can be written as a rank-one perturbation of a diagonal matrix,
\[
A=D+\bfa e\,\bfa u^{T},
\]
where
\[
D=\mathrm{diag}(-\mu_1,-\mu_2,\ldots,-\mu_n),\qquad
\bfa e=\beta
\begin{bmatrix}
p_1\\ p_2\\ \vdots\\ p_n
\end{bmatrix},
\qquad
\bfa u^{T}=
\begin{bmatrix}
\mu_1 & \mu_2 & \cdots & \mu_n
\end{bmatrix}.
\]
Hence, by the matrix determinant lemma,
\[
\det(A-\lambda I)
=
\det(D-\lambda I)\Bigl(1+\bfa u^{T}(D-\lambda I)^{-1}\bfa e\Bigr),
\]
and therefore the characteristic polynomial can be written as
\[
P_A(\lambda)
=
(-1)^n\prod_{i=1}^n(\mu_i+\lambda)
\left(
1-\beta\sum_{i=1}^n\frac{p_i\mu_i}{\mu_i+\lambda}
\right).
\]
As $\lambda \neq -\mu_{i}$ for any $i$, it follows that any eigenvalue is a root of
\[
q(\lambda):=
1-\beta\sum_{i=1}^n\frac{p_i\mu_i}{\mu_i+\lambda}.
\]
To locate the eigenvalues, first observe that on each interval
\[
(-\mu_i,-\mu_{i+1}),\qquad i=1,\ldots,n-1,
\]
the function $q$ is continuous and strictly increasing, since
\[
q'(\lambda)
=
\beta\sum_{i=1}^n\frac{p_i\mu_i}{(\mu_i+\lambda)^2}>0.
\]
Moreover,
\[
\lim_{\lambda\to-\mu_i^{+}} q(\lambda)=-\infty,
\qquad
\lim_{\lambda\to-\mu_{i+1}^{-}} q(\lambda)=+\infty.
\]
Hence $q$ has exactly one root in each interval $(-\mu_i,-\mu_{i+1})$, which gives
\[
-\mu_1 < \lambda_1 < -\mu_2 < \lambda_2 < \cdots < -\mu_{n-1} < \lambda_{n-1} < -\mu_n.
\]
It remains to locate the dominant eigenvalue $\lambda_n$, which is the unique root of $q$ in the interval $(-\mu_n,\infty)$. To do this, define
\[
h(\lambda):=\sum_{i=1}^n \frac{p_i\mu_i}{\mu_i+\lambda},
\qquad \lambda>-\mu_n.
\]
Then $q(\lambda)=1-\beta h(\lambda)$, and
\[
h'(\lambda)
=
-\sum_{i=1}^n \frac{p_i\mu_i}{(\mu_i+\lambda)^2}<0,
\]
so $h$ is strictly decreasing on $(-\mu_n,\infty)$. Since
\[
\lim_{\lambda\to-\mu_n^{+}} h(\lambda)=+\infty,
\qquad
\lim_{\lambda\to\infty} h(\lambda)=0,
\]
the equation
\[
h(\lambda)=\frac{1}{\beta}
\]
has a unique solution $\lambda_n$. We now consider the three cases separately.

\medskip
\noindent
\textbf{Case 1: $1<\beta\le 2$.}
Since $h(0)=1$ and $1/\beta<1$, it follows that $\lambda_n>0$. To obtain a lower bound, evaluate $h$ at $\lambda=(\beta-1)\mu_n$:
\[
h((\beta-1)\mu_n)
=
\sum_{i=1}^n\frac{p_i\mu_i}{\mu_i+(\beta-1)\mu_n}.
\]
Because $\mu_n\le \mu_i$ for all $i$,
\[
\mu_i+(\beta-1)\mu_n \le \mu_i+(\beta-1)\mu_i=\beta\mu_i,
\]
and hence
\[
h((\beta-1)\mu_n)
\ge
\sum_{i=1}^n \frac{p_i}{\beta}
=
\frac{1}{\beta}.
\]
Since $h$ is strictly decreasing, this implies
\[
(\beta-1)\mu_n < \lambda_n.
\]
To obtain the upper bound we first write 
\[
h(\lambda)
=
\sum_{i=1}^n\frac{p_i\mu_i}{\mu_i+\lambda}
=
1-\lambda\sum_{i=1}^n\frac{p_i}{\mu_i+\lambda}.
\]
Since the function $x\mapsto 1/(x+\lambda)$ is strictly convex 
for $\lambda>0$, Jensen's inequality gives
\[
\sum_{i=1}^n\frac{p_i}{\mu_i+\lambda}
>
\frac{1}{\bar{\mu}+\lambda}.
\]
We now let $\lambda=(\beta-1)\bar{\mu}$ and evaluate 
\[
h((\beta-1)\bar{\mu})
<
1-\frac{(\beta-1)\bar{\mu}}{\beta\bar{\mu}}
=
\frac{1}{\beta},
\]
and again using monotonicity of $h$ we obtain
\[
\lambda_n<(\beta-1)\bar{\mu}.
\]
Hence
\[
0<(\beta-1)\mu_n<\lambda_n<(\beta-1)\bar{\mu}.
\]

\medskip
\noindent
\textbf{Case 2: $\beta=1$.}
In this case
\[
q(0)=1-\sum_{i=1}^n p_i = 0,
\]
so $\lambda_n=0$. Since the remaining $n-1$ eigenvalues lie in the intervals
\[
(-\mu_1,-\mu_2),\ldots,(-\mu_{n-1},-\mu_n),
\]
they are all negative.

\medskip
\noindent
\textbf{Case 3: $0<\beta<1$.}
Now $1/\beta>1$, while $h(0)=1$, so the unique root of $h(\lambda)=1/\beta$ lies in the 
interval $(-\mu_n,0)$, and therefore $\lambda_n<0$. To obtain the upper bound, evaluate $h$ at $\lambda=(\beta-1)\mu_n<0$:
\[
h((\beta-1)\mu_n)
=
\sum_{i=1}^n\frac{p_i\mu_i}{\mu_i+(\beta-1)\mu_n}.
\]
Since $\mu_n\le \mu_i$,
\[
\mu_i+(\beta-1)\mu_n
=
\mu_i-(1-\beta)\mu_n
\ge
\mu_i-(1-\beta)\mu_i
=
\beta\mu_i,
\]
and hence
\[
h((\beta-1)\mu_n)
\le
\sum_{i=1}^n\frac{p_i}{\beta}
=
\frac{1}{\beta}.
\]
Since $h$ is strictly decreasing, it follows that
\[
\lambda_n<(\beta-1)\mu_n.
\]
For the lower bound, consider $\lambda=(\beta-1)\bar{\mu}$. If $(\beta-1)\bar{\mu}\le -\mu_n$, then the bound is immediate because
\[
\lambda_n>-\mu_n\ge (\beta-1)\bar{\mu}.
\]
If $(\beta-1)\bar{\mu}>-\mu_n$, then $h(\lambda)$ remains strictly convex since 
$\lambda>-\mu_{n}$ and $x\geq \mu_{n}$. Evaluating at $\lambda=(\beta-1)\bar{\mu}<0$, 
and using Jensen's inequality, we obtain
\[
h((\beta-1)\bar{\mu})
>
1-\frac{(\beta-1)\bar{\mu}}{\beta\bar{\mu}}
=
\frac{1}{\beta},
\]
and therefore
\[
(\beta-1)\bar{\mu}<\lambda_n.
\]
We have therefore established that
\[
{\rm max}[-\mu_{n},(\beta-1)\bar{\mu}] < \lambda_n < (\beta-1)\mu_n<0
\]
and this completes the proof.
\end{proof}

\subsection{Example}
\begin{figure}[h!]
\begin{center}
 \leavevmode
 \mbox{
\begin{minipage}{0.95\textwidth}
\begin{center}
\scalebox{0.55}{\includegraphics{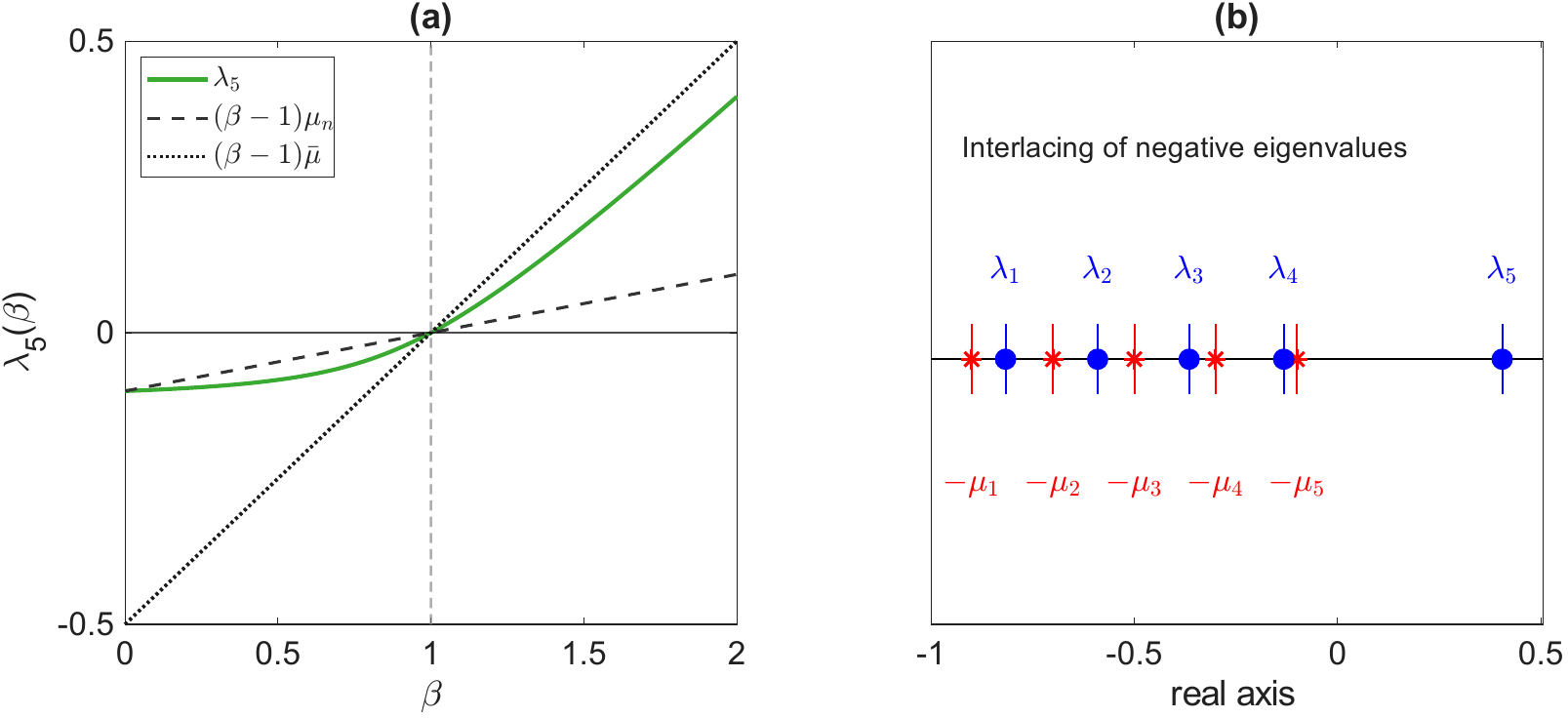}}
\end{center}
 \end{minipage}
}
\end{center}
\caption{
Spectral properties of the system matrix $A$ for an example with $n=5$. 
(a) Dependence of the dominant eigenvalue $\lambda_n$ on the stress parameter $\beta \in [0,2]$. 
The horizontal line $\lambda=0$ and the vertical line $\beta=1$ separate the decay ($\beta<1$) and growth ($\beta>1$) regimes. The dashed and dotted curves indicate the bounds $(\beta-1)\mu_n$ and $(\beta-1)\bar{\mu}$, respectively, between which $\lambda_n$ lies, in agreement with Theorem~\ref{th:3p}. 
(b) Real-line representation of the spectrum for $\beta=2$, showing the interlacing of the eigenvalues $\lambda_i$ with the points $-\mu_i$. In particular, one eigenvalue lies in each interval $(-\mu_i,-\mu_{i+1})$, illustrating the spectral structure established in Theorem~\ref{th:3p}.
}
\label{fig:eigs_npheno}
\end{figure}

As an illustrative example let's assume that the unit interval ${\cal V}=[0,1]$ is partitioned uniformly into $n$ intervals and that the division rates are the midpoints of each interval so that 
\[
\mu_{n}=\frac{1}{2n}, \quad {\rm and} \quad \mu_{i-1}=\mu_{i}+\frac{1}{n}, \quad i=n,\ldots,2.
\]
For this example, 
let us also assume a uniform probability that cells will be born into each phenotype so 
that $p_{i}=1/n$, $i=1,\ldots,n$. To illustrate the spectral properties of the $n$-phenotype 
system, we consider a simple example with $n=5$ and examine the dependence of the eigenvalues 
of the system matrix $A$ on the stress parameter $\beta$. The eigenvalues are plotted 
in Figure~\ref{fig:eigs_npheno}. Panel (a) shows the dominant eigenvalue $\lambda_5(\beta)$ as 
functions of $\beta \in [0,2]$. This representation highlights the transition 
at $\beta=1$, where the dominant eigenvalue changes sign, separating growth and decay regimes. 
In particular, $\lambda_5$ becomes positive for $\beta>1$. This behaviour is consistent 
with the bounds established in Theorem~\ref{th:3p}. Panel (b) presents the spectrum for a 
fixed value $\beta=2$ on the real line, together with the points $-\mu_i$. 
This provides a clear visual illustration of the interlacing property of the 
non-dominant eigenvalues, as described in Theorem~\ref{th:3p}, with each eigenvalue 
lying strictly between successive values of $-\mu_i$. Taken together, these plots demonstrate both the dependence of the dominant eigenvalue 
on the stress parameter and the underlying spectral structure of the system.

\section{Continuous phenotype model}

The $n$-phenotype formulation considered in Section~3 naturally suggests a 
limiting description in which the division rates are distributed over a continuum. 
We therefore consider a continuous phenotype model in which the division rate 
$\mu$ is treated as a continuous variable. Biologically, a cell's division rate 
$\mu$ is restricted by the kinetic limitations of the cell cycle, precluding 
an infinitely fast replication cycle or an infinitely slow one. Consequently, 
we define our division rate space over a compact, biologically constrained 
interval \(\mathcal{V} = [\mu_{\min}, \mu_{\max}]\). Let $N(\mu,t)$ denote the density of cells with division rate $\mu \in \mathcal{V}$ at time $t$. 
Upon division, each daughter cell is assigned a division rate according to a 
strictly positive probability density $p(\mu)$, with $p(\mu) > 0$ for all 
$\mu \in \mathcal{V}$ and
\[
\int_{\mathcal V} p(\mu)\, d\mu = 1.
\]
As in Sections~2 and 3, environmental stress acts by reducing viability at birth, 
characterised by the factor $\beta=2(1-\sigma)$. Under these assumptions, the population dynamics are governed by the integro-differential equation
\begin{equation}
\frac{\partial N(\mu,t)}{\partial t}
=
\beta p(\mu)\int_{\mathcal V} \nu N(\nu,t)\, d\nu - \mu N(\mu,t).
\label{eq:continuous_model}
\end{equation}
We define the mean division rate as
\[
\bar{\mu} = \int_{\mathcal V} \mu\, p(\mu)\, d\mu.
\]

\subsection{Spectral structure and asymptotic behaviour}
Equation~\eqref{eq:continuous_model} defines a linear operator $\mathcal{A}$ on the Banach space $C(\mathcal{V})$, whose action is given by $\mathcal{A} f(\mu) = \beta p(\mu)\int_{\mathcal V} \nu f(\nu)\, d\nu - \mu f(\mu)$. The long-term behavior of the population is governed by the spectrum of $\mathcal{A}$, which can be fully characterized by determining the values of $\lambda \in \mathbb{C}$ for which the resolvent operator $R(\lambda, \mathcal{A}) = (\lambda I - \mathcal{A})^{-1}$ fails to exist as a bounded operator.

\begin{theorem}
\label{th:continuous}
The spectrum $\sigma(\mathcal{A})$ consists of a continuous spectrum $\sigma_c(\mathcal{A}) = [-\mu_{\max}, -\mu_{\min}]$ and a unique, isolated real eigenvalue $\lambda_0 \in \mathbb{R} \setminus \sigma_c(\mathcal{A})$. This dominant eigenvalue $\lambda_0$ is the unique real root of the dispersion relation
\begin{equation}
F(\lambda;p) = 1 - \beta \int_{\mathcal V} \frac{\mu p(\mu)}{\mu + \lambda}\, d\mu = 0.
\label{eq:continuous_root}
\end{equation}
Moreover, this dominant eigenvalue satisfies
\[
0 < \lambda_0 < (\beta-1)\bar{\mu},
\qquad 1<\beta \le 2,
\]
\[
\lambda_0 = 0,
\qquad \beta=1,
\]
and
\[
\max[-\mu_{\min},(\beta-1)\bar{\mu}] < \lambda_0 < 0,
\qquad 0<\beta<1.
\]
\end{theorem}

\begin{proof}
To establish the spectrum via the resolvent operator, we consider the inhomogeneous operator equation $(\lambda I - \mathcal{A})f = g$ for an arbitrary target function $g(\mu) \in C(\mathcal{V})$. Expanding this expression using the definition of $\mathcal{A}$ yields
\[
(\lambda + \mu)f(\mu) - \beta p(\mu)\int_{\mathcal V} \nu f(\nu)\, d\nu = g(\mu).
\]
For any $\lambda$ outside the real interval $[-\mu_{\max}, -\mu_{\min}]$, the term $(\lambda + \mu)$ is non-zero across the entire domain. We can therefore divide by this term and isolate $f(\mu)$:
\begin{equation}
f(\mu) = \frac{g(\mu)}{\lambda + \mu} + \frac{\beta p(\mu)}{\lambda + \mu}\int_{\mathcal V} \nu f(\nu)\, d\nu.
\label{eq:resolvent_step1}
\end{equation}
The integral on the right-hand side evaluates to a scalar constant, which we define as $C = \int_{\mathcal V} \nu f(\nu)\, d\nu$. Multiplying~\eqref{eq:resolvent_step1} by $\mu$ and integrating over $\mathcal{V}$ allows us to self-consistently solve for $C$:
\[
C = \int_{\mathcal V} \frac{\mu g(\mu)}{\lambda + \mu}\, d\mu + C \beta \int_{\mathcal V} \frac{\mu p(\mu)}{\lambda + \mu}\, d\mu.
\]
Grouping the terms containing $C$ reveals the dispersion function $F(\lambda;p)$ as defined in~\eqref{eq:continuous_root}:
\[
C F(\lambda;p) = \int_{\mathcal V} \frac{\mu g(\mu)}{\lambda + \mu}\, d\mu.
\]
Substituting this back into~\eqref{eq:resolvent_step1} yields the explicit action of the resolvent operator $R(\lambda, \mathcal{A})g = f$:
\begin{equation}
R(\lambda, \mathcal{A})g(\mu) = \frac{g(\mu)}{\lambda + \mu} + \frac{\beta p(\mu)}{\lambda + \mu} \cdot \frac{\int_{\mathcal V} \frac{\nu g(\nu)}{\lambda + \nu}\, d\nu}{F(\lambda;p)}.
\label{eq:explicit_resolvent}
\end{equation}
From this explicit form, the breakdown of the resolvent reveals the two components of the spectrum:
\begin{enumerate}
    \item \textbf{Continuous Spectrum:} If $\lambda \in [-\mu_{\max}, -\mu_{\min}]$, the local denominators $(\lambda + \mu)$ vanish point-by-point, making the operator unboundedly singular over a continuous interval. This defines the continuous spectrum $\sigma_c(\mathcal{A}) = [-\mu_{\max}, -\mu_{\min}]$.
    \item \textbf{Point Spectrum (Isolated Eigenvalue):} Outside of this continuous interval, a singularity occurs if and only if the global denominator vanishes, which yields the dispersion relation $F(\lambda_0;p) = 0$.
\end{enumerate}
To establish the existence, uniqueness, and bounds of the root $\lambda_0$, we analyze the properties of $F(\lambda;p)$ on the open domain $\lambda \in (-\mu_{\min}, \infty)$. Differentiating with respect to $\lambda$ gives
\[
\frac{\partial F}{\partial \lambda} = \beta \int_{\mathcal V} \frac{\mu p(\mu)}{(\mu + \lambda)^2}\, d\mu > 0.
\]
Because the derivative is strictly positive, $F(\lambda;p)$ is strictly monotonic and increasing on this domain. Evaluating the function at the origin yields $F(0;p) = 1 - \beta \int_{\mathcal V} p(\mu)\, d\mu = 1 - \beta$, while taking the asymptotic limit yields $\lim_{\lambda \to \infty} F(\lambda;p) = 1$. 

We now consider the three cases for the viability parameter $\beta$:
\begin{itemize}
    \item For $1 < \beta \le 2$, we have $F(0;p) = 1 - \beta < 0$. Since $\lim_{\lambda \to \infty} F(\lambda;p) = 1 > 0$, the Intermediate Value Theorem combined with strict monotonicity guarantees a unique, strictly positive real root $\lambda_0 > 0$. 
    \item For $\beta = 1$, evaluating the dispersion relation at zero gives $F(0;p) = 1 - 1 = 0$, making $\lambda_0 = 0$ the unique real root.
    \item For $0 < \beta < 1$, we have $F(0;p) = 1 - \beta > 0$. As $\lambda \to -\mu_{\min}^+$ from the right, the singular denominator causes the integral to blow up to infinity, meaning $\lim_{\lambda \to -\mu_{\min}^+} F(\lambda;p) = -\infty$. This guarantees a unique, strictly negative real root trapped in the gap $\lambda_0 \in (-\mu_{\min}, 0)$.
\end{itemize}
The corresponding upper bounds for $1 < \beta \le 2$, as well as the lower bounds $(\beta - 1)\bar{\mu}$, are established through algebraic monotonicity checks of the integral expression in a manner identical to the discrete case detailed in Section~3.
\end{proof}

\subsection{Asymptotic population structure}

The dominant eigenfunction associated with~\eqref{eq:continuous_model} determines the asymptotic composition of the population. In particular, writing the solution in the form
\[
N(\mu,t) \sim C e^{\lambda_0 t} \phi(\mu),
\]
the function $\phi(\mu)$ represents the long-term distribution of phenotypes. From the form of the eigenfunction,
\[
\phi(\mu) \propto \frac{p(\mu)}{\mu+\lambda_0},
\]
it follows that the asymptotic population is biased towards lower values of $\mu$. 
In particular, under increasing levels of stress (i.e. decreasing $\beta$), 
the dominant eigenvalue $\lambda_0$ decreases, leading to a stronger weighting towards 
slower-dividing phenotypes. This behaviour is consistent with the numerical observations in 
Sections~2 and 3, and reflects the same selection mechanism: environmental stress 
suppresses faster-dividing cells, resulting in a population that is asymptotically 
dominated by slower-dividing phenotypes. Thus, the continuous model confirms that the influence of heterogeneity on both the growth rate and the population structure is preserved in the limit of infinitely many phenotypes.

\subsection{Examples}
As a simple illustrative baseline, suppose $\mathcal V=[0,1]$, $p(\mu)=1$, and $\beta=2$, corresponding to a uniform distribution of division rates in the absence of environmental stress. In this case, (\ref{eq:continuous_root}) reduces to
\[
\int_0^1 \frac{\mu}{\lambda+\mu}\,d\mu=\frac12,
\]
which can be integrated analytically to yield the transcendental relationship
\begin{equation}
\lambda \ln\!\left(1+\frac{1}{\lambda}\right)=\frac12.
\label{eq:lambda_uniform}
\end{equation}
Solving \eqref{eq:lambda_uniform} numerically gives $\lambda \approx 0.398$, while the mean division 
rate is $\bar{\mu}=\int_0^1 \mu\,d\mu=0.5$. Therefore, the asymptotic growth rate of the heterogeneous 
continuous model is strictly smaller than the growth rate predicted by the corresponding homogeneous model 
based on the mean division rate. 

To demonstrate that this framework remains analytically tractable under non-uniform phenotypic profiles, we also consider a skewed power-law allocation at birth, $p(\mu) = 2\mu$ on $\mathcal{V}=[0,1]$ with $\beta=2$. This profile shifts the biological weight toward faster-dividing phenotypes, yielding a higher mean division rate of $\bar{\mu} = \frac{2}{3}$. For this distribution, equation \eqref{eq:continuous_root} integrates to
\begin{equation}
1 = 4 \left( \frac{1}{2} - \lambda + \lambda^2 \ln\left(1 + \frac{1}{\lambda}\right) \right).
\label{eq:lambda_skewed}
\end{equation}
Solving \eqref{eq:lambda_skewed} yields an asymptotic growth rate of $\lambda \approx 0.616$. As observed in the uniform baseline, the long-term growth rate remains systematically lower than the homogeneous expectation ($\lambda < \bar{\mu}$), proving that the selective suppression of fast-dividing cells is structurally robust to the underlying shape of the allocation function.

To illustrate the transient dynamics, consider the initial condition where the population profile reflects its birth distribution:
\begin{equation}
N(\mu,0)=p(\mu), \qquad \mu\in\mathcal V.
\label{eq:initcont}
\end{equation}
The initial rate of change of the total population is then
\[
\left.\frac{d}{dt}\int_{\mathcal V}N(\mu,t)\,d\mu\right|_{t=0}
=
\int_{\mathcal V}\mu\,N(\mu,0)\,d\mu
=
\bar{\mu},
\]
showing that the population initially expands at the homogeneous mean rate ($1/2$ for the uniform baseline and $2/3$ for the skewed profile). 
However, as $t\to\infty$, the phenotypic composition shifts and the growth rate converges 
to the smaller value $\lambda$ determined by (\ref{eq:continuous_root}).

\begin{figure}[t!]
\centering
\includegraphics[width=\textwidth]{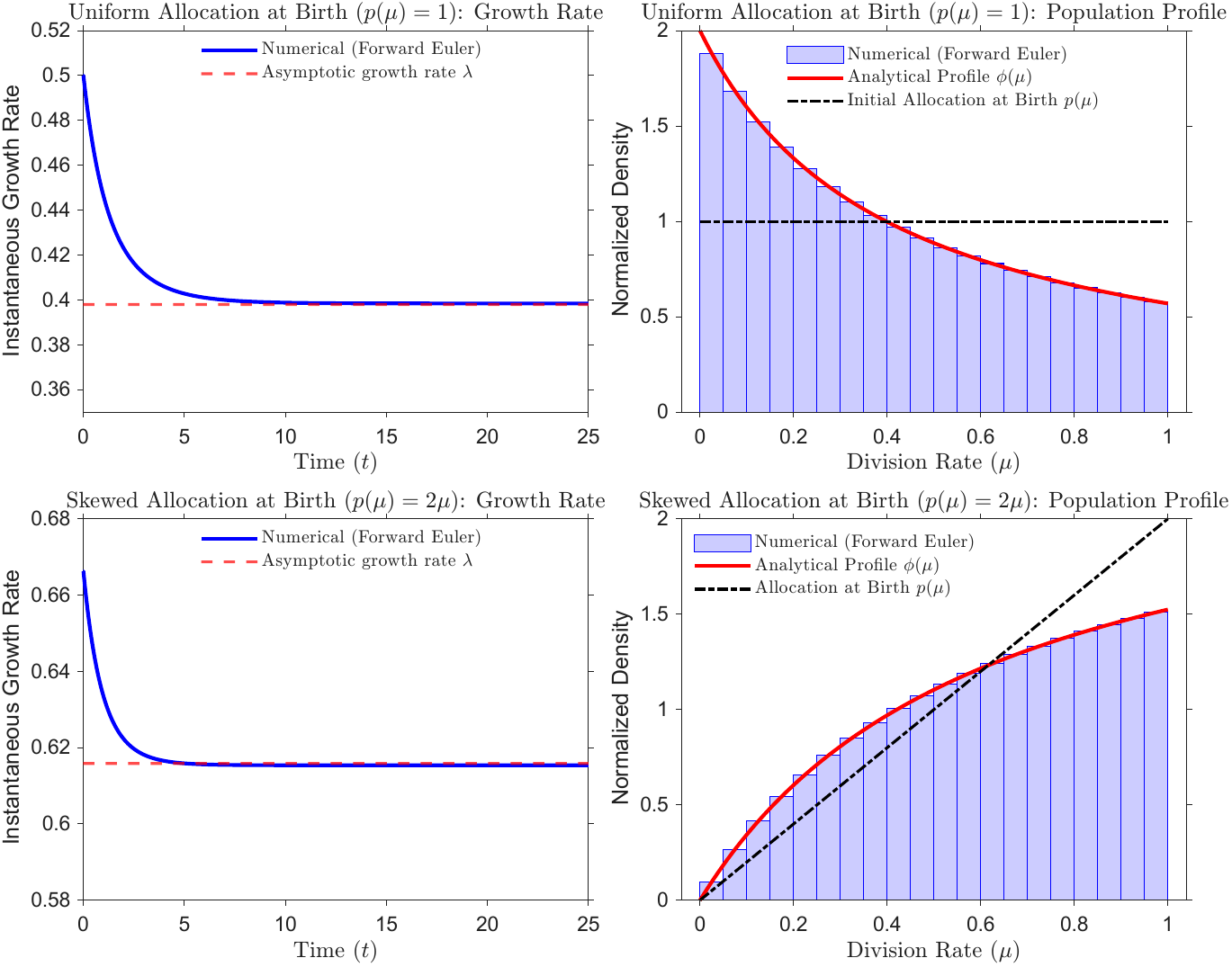} 
\caption{Transient growth rates and asymptotic phenotypic distributions ($\beta = 2$). Left panels (a, c) show the instantaneous population growth rate (solid blue lines) converging onto the analytical dominant eigenvalue $\lambda$ (dashed red lines). Right panels (b, d) show the normalized asymptotic phenotypic distribution (blue bars for numerical integration, solid red lines for analytical predictions) overlaid against the allocation at birth $p(\mu)$ (black dash-dotted lines). The top row (a, b) assumes a uniform allocation profile $p(\mu) = 1$, and the bottom row (c, d) assumes a skewed power-law allocation profile $p(\mu) = 2\mu$.}
\label{fig:const1}
\end{figure}

We compute numerical solutions of \ref{eq:continuous_model} using a uniform partition of 
$\mathcal V=[0,1]$ into $M=20$ subintervals and a forward Euler time-stepping scheme. 
Figure~\ref{fig:const1} demonstrates these transient and asymptotic dynamics for both allocation profiles $p(\mu)$. 
As shown in Figures~\ref{fig:const1}(a) and (c), the instantaneous growth rates begin precisely at their 
respective homogeneous mean rates $\bar{\mu}$ and decrease monotonically over time, converging towards
the analytical values $\lambda$ guaranteed by Theorem \ref{th:continuous}. 
Figures~\ref{fig:const1}(b) and (d) depict the normalised asymptotic population profiles as 
a function of $\mu$, plotted alongside the analytical predictions derived from the 
dominant eigenfunction $\phi(\mu) \propto \frac{p(\mu)}{\lambda+\mu}$. 
The numerical and analytical profiles are in excellent agreement for both cases. 

Overlaying these steady-state densities directly against the allocation at birth $p(\mu)$ visually 
highlights the intensity of the selection mechanism. In the uniform case, the profile shifts to 
over-represent slow-dividing states. In the skewed case—where the allocation at birth distribution 
is heavily biased toward the fastest replication rate ($\mu=1$)—the continuous evolutionary
process significantly reduces the contribution from the high-division boundary. 
Therefore, the continuous model confirms that the influence of heterogeneity on both the 
growth rate and the population structure is preserved in the limit of infinitely many phenotypes.

\subsection{Dependence of the growth rate on the division-rate distribution}

In Section~\ref{sec:cv}, we established monotonicity results for the growth rate in the discrete
two-phenotype model, showing how the growth rate varies with the coefficient of variation of the
division rates when the mean division rate is held fixed. A natural question is whether an
analogous result holds in the continuous setting. Specifically, we wish to determine how the
asymptotic growth rate \(\lambda_0\) depends on the shape of the distribution \(p(\mu)\) of
division rates within the population. To address this question, we consider a family of distributions related through
\emph{mean-preserving spreads} (see, e.g., \cite{Rothschild1970} or \cite{Shaked2007}). 
Let \(p_0(\mu)\) be a baseline distribution with mean
\(\bar{\mu}\). A distribution \(p_1(\mu)\) is said to be a mean-preserving spread of \(p_0(\mu)\)
if
\begin{enumerate}
    \item \(p_1(\mu)\) and \(p_0(\mu)\) have the same mean \(\bar{\mu}\);
    \item \(p_1(\mu)\) is more dispersed than \(p_0(\mu)\) in the convex-order sense, i.e.
    \[
        \mathbb{E}_{p_1}[f(\mu)]
        \ge
        \mathbb{E}_{p_0}[f(\mu)]
    \]
    for every convex function \(f\).
\end{enumerate}
The second condition implies that \(p_1\) places relatively more weight on both low and high
division rates while preserving the population mean.

\begin{theorem}
\label{thm:mps}   
Let \(p_1(\mu)\) be a mean-preserving spread of \(p_0(\mu)\). If
\(\lambda_0\) and \(\lambda_1\) denote the corresponding unique, isolated eigenvalues guaranteed by Theorem~\ref{th:continuous}, then
\begin{align}
0 < \lambda_1 < \lambda_0, &\qquad 1 < \beta \le 2,\\
\lambda_1=\lambda_0=0, &\qquad \beta =1,\\
\max[-\mu_{\min},(\beta-1)\bar{\mu}] < \lambda_0 < \lambda_1 < 0, &\qquad 0<\beta<1.
\end{align}
\end{theorem}

\begin{proof}
Let $F(\lambda;p)$ be the dispersion function defined in~\eqref{eq:continuous_root}, written in expectation form as
\[
F(\lambda;p) = 1-\beta \mathbb{E}_{p}\!\left[f(\mu;\lambda)\right], \quad \text{where} \quad f(\mu;\lambda)=\frac{\mu}{\mu+\lambda}.
\]
By Theorem~\ref{th:continuous}, $F(\lambda;p)$ is strictly increasing in $\lambda$ on the domain $(-\mu_{\min}, \infty)$ for any admissible distribution $p$. To evaluate the shape of the integrand with respect to $\mu$, we compute its second derivative:
\[
\frac{\partial^2 f}{\partial \mu^2} = -\frac{2\lambda}{(\mu+\lambda)^3}.
\]

We evaluate the three cases for the viability parameter $\beta$:
\begin{itemize}
    \item \textbf{Case $1 < \beta \le 2$:} From Theorem~\ref{th:continuous}, the roots satisfy $\lambda_0, \lambda_1 > 0$. For any $\lambda > 0$, the second derivative $\partial^2 f/\partial \mu^2 < 0$, making $f(\mu;\lambda)$ strictly concave in $\mu$. Because $p_1$ is a mean-preserving spread of $p_0$, the definition of convex ordering yields $\mathbb{E}_{p_1}[f(\mu;\lambda_0)] < \mathbb{E}_{p_0}[f(\mu;\lambda_0)]$. Evaluating $F(\lambda;p_1)$ at the baseline root $\lambda = \lambda_0$ gives:
    \[
    F(\lambda_0;p_1) = 1-\beta \mathbb{E}_{p_1}\!\left[f(\mu;\lambda_0)\right] > 1-\beta \mathbb{E}_{p_0}\!\left[f(\mu;\lambda_0)\right] = F(\lambda_0;p_0) = 0.
    \]
    Since $F(\lambda;p_1)$ is strictly increasing in $\lambda$, its unique zero must lie to the left of $\lambda_0$, establishing that $\lambda_1  <\lambda_0$. 
  \item \textbf{Case $\beta=1$:} For any viable distribution $p$, 
  the unique root is $\lambda=0$, so $\lambda_1 = \lambda_0 = 0$.
\item \textbf{Case $0 < \beta < 1$:} From Theorem~\ref{th:continuous}, 
the roots satisfy $-\mu_{\min}<\lambda_0, \lambda_1 < 0$. For any $-\mu_{\min}<\lambda < 0$, and $\mu \in \cal{V}$, the 
second derivative $\partial^2 f/\partial \mu^2 > 0$, making $f(\mu;\lambda)$ strictly convex in $\mu$. The convex ordering inequality is reversed, giving $\mathbb{E}_{p_1}[f(\mu;\lambda_0)] > \mathbb{E}_{p_0}[f(\mu;\lambda_0)]$. Evaluating at $\lambda = \lambda_0$ yields:
    \[
    F(\lambda_0;p_1) = 1-\beta \mathbb{E}_{p_1}\!\left[f(\mu;\lambda_0)\right] < 1-\beta \mathbb{E}_{p_0}\!\left[f(\mu;\lambda_0)\right] = F(\lambda_0;p_0) = 0.
    \]
    Because $F(\lambda;p_1)$ is strictly increasing, its unique zero must lie to the right of $\lambda_0$, establishing that $\lambda_1 > \lambda_0$.
\end{itemize}
\end{proof}

\begin{figure}[htbp]
    \centering
    \begin{subfigure}[b]{0.48\textwidth}
        \centering
        \includegraphics[width=\textwidth]{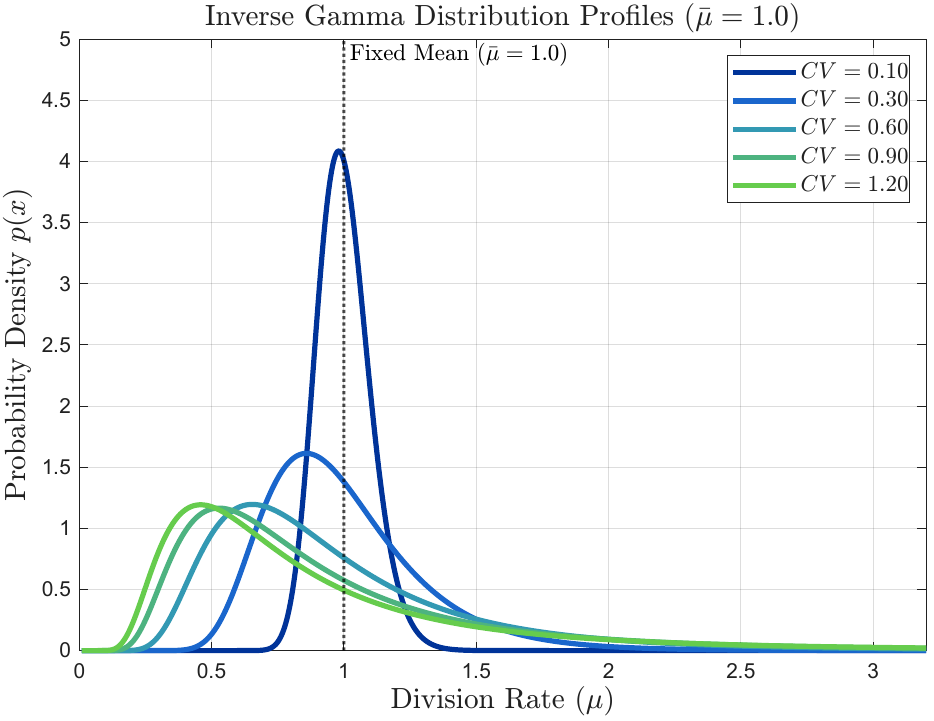} 
        \caption{Sample mean-preserving spread distributions of the inverse-gamma distribution with fixed mean $\bar{\mu} = 1$ and varying coefficient of variation (CV).}
        \label{fig:plot1}
    \end{subfigure}
    \hfill 
    \begin{subfigure}[b]{0.48\textwidth}
        \centering
        \includegraphics[width=\textwidth]{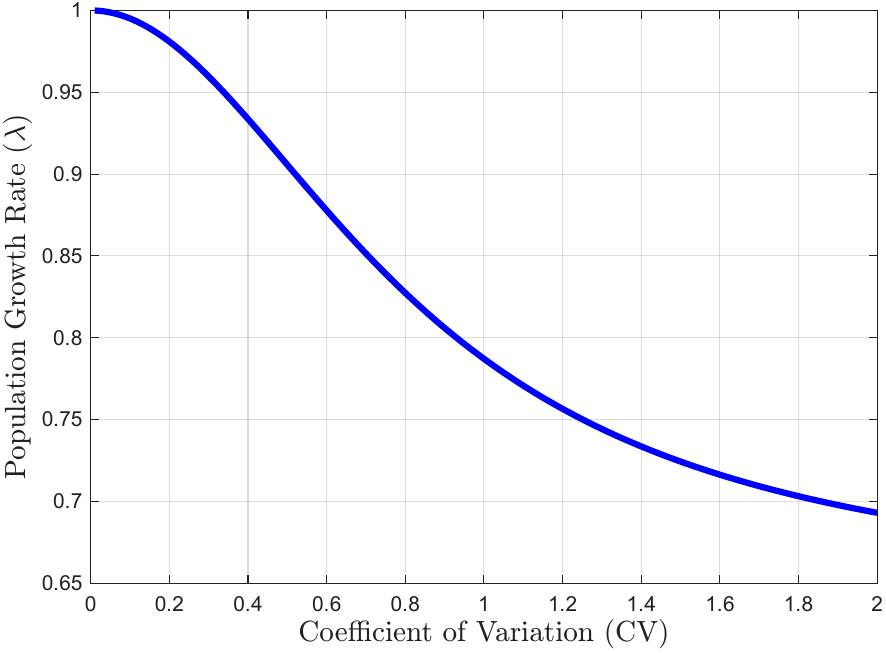}
        \caption{Population growth rate $\lambda$ when $\beta=2$ as a function of the coefficient of variation (CV) of the inverse-gamma distribution.}
        \label{fig:plot2}
    \end{subfigure}
    \caption{Illustration of the effect of increasing variability in division
rates on population growth. (a) Inverse-gamma distributions with fixed mean
$\bar{\mu}=1$ and increasing coefficient of variation, forming a family of
mean-preserving spreads. (b) Corresponding asymptotic growth rates when $\beta=2$. As predicted by Theorem~\ref{thm:mps},
increased dispersion in division rates reduces the growth rate when
$\beta>1$.}
    \label{fig:both_plots}
\end{figure}

To illustrate Theorem~\ref{thm:mps}, we consider a family of inverse-gamma
division-rate distributions with fixed mean division rate. This choice is
natural because experimental studies often characterise variability in terms
of cell-cycle durations rather than division rates. If the division time
\(T\) is gamma distributed, then the corresponding division rate
\(\mu=1/T\) follows an inverse-gamma distribution. The inverse-gamma family
therefore provides a biologically relevant class of division-rate
distributions whose variability can be varied systematically while
preserving the mean. Figure~\ref{fig:both_plots}(a) shows several inverse-gamma distributions with
mean \(\bar{\mu}=1\) and increasing coefficient of variation, forming a
sequence of mean-preserving spreads. Figure~\ref{fig:both_plots}(b) shows the
corresponding asymptotic growth rates for \(\beta=2\). Consistent with
Theorem~\ref{thm:mps}, increasing the variability in division rates leads to
a monotonic decrease in the population growth rate.

\section{Biological significance}
\label{sec:bio}

\begin{figure}[h!]
\centering
\begin{minipage}{0.48\textwidth}
  \centering
  \scalebox{1.1}{\includegraphics{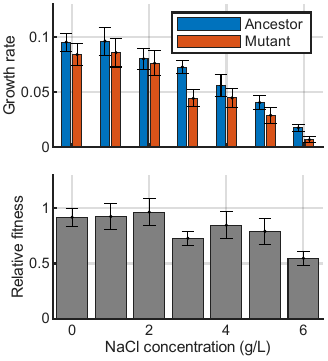}}\\ 
  {\small (a) Genotype CC-2344.}
\end{minipage}\hfill
\begin{minipage}{0.48\textwidth}
  \centering
  \scalebox{1.1}{\includegraphics{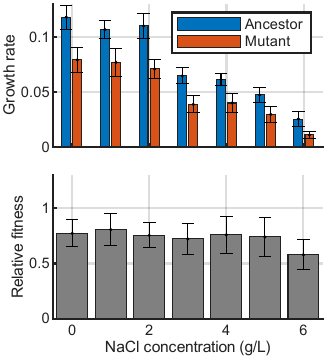}}\\
  {\small (b) Genotype CC-2931.}
\end{minipage}
\caption{Mean population growth rates ($\bar{r}$, per hour) averaged over experimental 
replicates of mutation accumulation \emph{C. reinhardtii} genotypes and their 
respective ancestors (genotypes CC-2344 and CC-2931) under seven different 
levels of stress (expressed as concentration of NaCl in the environment). 
Bottom panels show relative fitnesses calculated as in \eqref{eq:rf}. 
Data from \cite{kraemer2016}.}
\label{fig:data}
\end{figure}

Here we illustrate how the models analysed in the previous sections can 
be used to generate new interpretations of biological data. The datasets 
displayed in Figure~\ref{fig:data} were obtained from \cite{kraemer2016} 
and represent the sample mean population growth rates (per hour) evaluated over 
experimental replications of distinct genotypes of the green alga 
\emph{Chlamydomonas reinhardtii}, a widely used model organism in ecology 
and evolutionary biology. For each of the two genotypes (CC-2344 and CC-2931), the blue bars show the 
mean empirical population growth rates of the ancestral strain ($\bar{r}_a$), while 
the red bars show the corresponding mean population growth rates across 15 mutation 
accumulation lines derived from that ancestor ($\bar{r}_m$). In these 
experiments, environmental stress is imposed by increasing the concentration 
of NaCl in the growth medium (horizontal axis). The bottom panels show the fitness of the mutants relative to the ancestor. 
When time is measured in units of ancestral generation time \cite{Chevin2011}, 
this relative fitness $w$ is given by:
\begin{equation} \label{eq:rf}
w = \exp\!\left( \frac{\bar{r}_m - \bar{r}_a}{\bar{r}_a}\ln{2} \right) \approx  1 + \frac{\bar{r}_m - \bar{r}_a}{\bar{r}_a}\ln{2}.
\end{equation}
These data suggest that mutational effects are deleterious to individuals and 
become more pronounced under increasing stress \cite{kraemer2016}. This dataset 
therefore provides a natural setting in which to explore how the 
density-independent mechanisms identified in this paper may offer an alternative 
interpretation of stress-dependent fitness effects.

\begin{figure}[h!]
\centering
\begin{minipage}{0.48\textwidth}
  \centering
  \scalebox{1.1}{\includegraphics{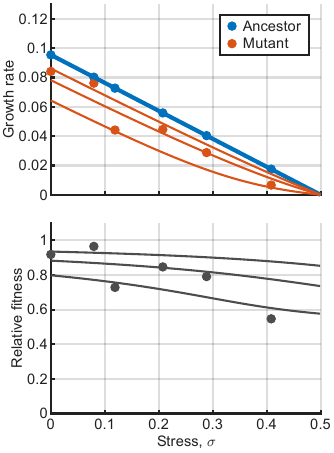}}\\
  {\small (a) Varying mutant $\mathrm{CV}$.}
\end{minipage}\hfill
\begin{minipage}{0.48\textwidth}
  \centering
  \scalebox{1.1}{\includegraphics{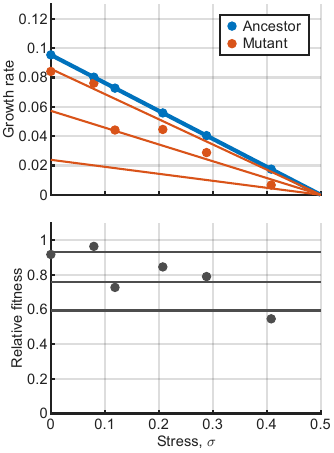}}\\
  {\small (b) Varying mutant $\bar{\mu}$.}
\end{minipage}
\caption{Model curves for varying mutant $\mathrm{CV}$ (a) and mutant $\bar{\mu}$ (b), 
superposed on CC-2344 data from Figure~\ref{fig:data}(a) generated by the 
2-phenotype model~\eqref{eq:ode2ph} assuming a discrete distribution with $p=0.7$. 
The curves correspond to the following mutant parameter values (from top to bottom): 
(a) $\mathrm{CV}=0.39$, $0.50$, $0.61$ (with $\bar{\mu}=0.095$); 
(b) $\bar{\mu}=0.084$, $0.057$, $0.024$ (with $\mathrm{CV}=0$). The ancestral 
genotype has fixed $\mathrm{CV}=0$ and $\bar{\mu}=0.095$.}
\label{fig:contours1}
\end{figure}

In Figure~\ref{fig:contours1} we overlay the outputs of the 2-phenotype 
model~\eqref{eq:ode2ph} with the experimental data for \emph{C. reinhardtii} 
genotype CC-2344. We begin by assuming that the ancestral population exhibits 
no intragenotypic variation in division rates, so that its population growth 
rate depends exclusively on the baseline division rate. In this case, the 
relationship between model growth rate $\lambda$ and stress $\sigma$ is linear, 
represented by the straight line joining the points $(0,\bar{\mu})$ and $(0.5,0)$ 
on the $(\sigma,\lambda)$-plane. To relate our model to the empirical setting, we map each NaCl concentration 
to a corresponding theoretical stress level $\sigma$ by requiring that the 
observed ancestral growth rates (blue points) align exactly with this linear 
relationship. This provides a direct mapping between NaCl concentration and 
$\sigma$ used throughout the remaining analysis. We then consider the mutation accumulation lines (red points) and compare two 
alternative biological mechanisms. In case (a), we fix the mean division 
rate $\bar{\mu}$ and introduce phenotypic variability by allowing 
$\mathrm{CV} > 0$. In case (b), we impose $\mathrm{CV} = 0$ and instead 
vary the underlying mean division rate. For each scenario, we calculate 
the asymptotic growth rate of the mutant according to model~\eqref{eq:ode2ph}, 
given by the dominant eigenvalue $\lambda^+$ in~\eqref{eq:evalues2ph}. The resulting model curves are shown in red, and the corresponding relative 
fitness curves (bottom panels) are computed using equation~\eqref{eq:rf}. This 
construction provides a direct comparison between the effects of phenotypic 
variability versus directional changes in mean division rate under varying stress, 
allowing us to assess different mechanisms for their capability to reproduce 
the observed amplification of relative fitness.

The two scenarios attempt to explain the data in meaningfully different ways. 
In case (a), mutation does not affect the mean of individual fitnesses 
(or division rates) but increases their variability. 
In case (b), intragenotypic variation is absent and mutation is taken 
to reduce the mean fitness of genotypes.

It is important to distinguish between variability arising within individual 
mutation lines and variability arising from differences in mean division rates between 
lines. The mutation accumulation dataset consists of multiple mutation lines, each 
characterised by a distribution of cellular division  rates. In the analysis above, the coefficient of variation represents average within-line variation (not variation between lines). 
This provides a mechanism by which the mutation accumulation population may effectively behave as a heterogeneous population in the sense of the model. 
Under a mean-only description, where each line is homogeneous with growth rate 
$(\beta-1)\mu$, differences between lines would not 
produce amplification of relative fitness under stress. By contrast, variability 
within lines introduces nonlinear effects on the growth rate through the 
eigenvalue structure of the model, leading to stress-dependent amplification. 
This suggests that increased within-genotype variability, rather than differences 
in mean fitness alone, may play a key role in explaining the data.

\begin{figure}[h!]
\begin{center}
 \leavevmode
 \mbox{
\begin{minipage}{0.5\textwidth}
\begin{center}
\scalebox{1.1}{\includegraphics{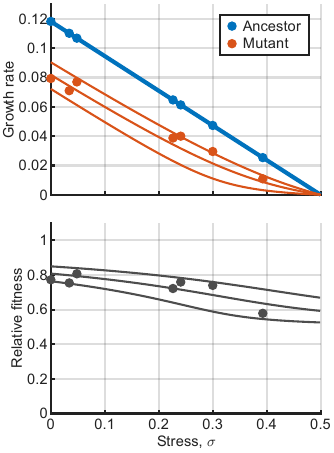}}\\ 
 {\small (a) Varying mutant CV.}
\end{center}
 \end{minipage}
\begin{minipage}{0.5\textwidth}
\begin{center}
\scalebox{1.1}{\includegraphics{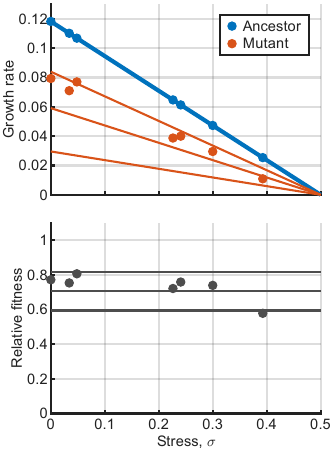}}\\ 
 {\small (b) Varying mutant $\overline{\mu}$.}
\end{center}
 \end{minipage}
}
\end{center}
 \caption{Model curves for varying mutant $\mathrm{CV}$ (a) 
 and mutant $\bar{\mu}$ (b), 
 superposed on CC-2931 
 data from Figure \ref{fig:data}(b) generated by the 
 2-phenotype model \eqref{eq:ode2ph} assuming a discrete distribution with $p=0.7$. The curves 
 correspond to the following mutant parameter values (from top to bottom): (a) $\mbox{CV}=0.55$, $0.60$, $0.64$ (with $\overline{\mu}=0.12$); (b) $\overline{\mu}=0.085$, $0.06$, $0.03$ (with $\mbox{CV}=0$). The ancestral genotype was assumed to have $CV=0$ and $\overline{\mu}=0.12$. }
\label{fig:contours2}
\end{figure}

The same procedure was applied to genotype CC-2931, for which a similar decline 
in relative fitness with stress was observed and reproduced by the 2-phenotype model (Figure~\ref{fig:contours2}). 
In both genotypes, the observed trends are consistent with a scenario where 
the ancestor exhibits lower variability in division rates, while the mutants 
display increased variability. While more detailed statistical 
analysis would be required to quantify the extent to which the data support this 
interpretation, the framework developed here provides a simple mechanistic 
explanation for the observed patterns and highlights the potential importance 
of variability in shaping population responses to environmental stress.

\section{Conclusions}
We analysed models originally developed by \cite{gomes2019effects}, where inter-individual 
variation in cell division rates was integrated into population growth dynamics. Our results 
demonstrate that this individual-level variation reduces population growth in favourable regimes, 
but conversely mitigates population decline in detrimental environments. Furthermore, by 
treating environmental stress as an explicit model parameter, we showed that the resulting 
reduction in population growth rates scales nonlinearly and becomes aggravated under stress.

We applied this theoretical framework to empirical data on growth rates from populations of 
the green alga \emph{C. reinhardtii}. Specifically, we examined the growth rates of two ancestral 
strains (CC-2344 and CC-2931) and their respective mutation accumulation (MA) lines along an 
environmental gradient of NaCl, a stressor known to impair \emph{C. reinhardtii} performance 
\cite{kraemer2016}. The empirical data reveal that MA lines exhibit consistently lower sample 
mean growth rates ($\bar{r}_m < \bar{r}_a$) than their ancestors, a deficit that amplifies with 
increasing stress. While the conventional interpretation of these trends is that individual mutants divide more 
slowly and suffer higher baseline vulnerability to stress, our models demonstrate that an 
entirely different biological mechanism is equally compatible with the data. Specifically, we 
propose that the observed patterns could stem from MA lines exhibiting greater individual-level 
variance in cell division rates. This variance effect alone is sufficient to explain the 
experimental trends. Even if the underlying mean division rates across individuals remained 
identical to (or higher than) the ancestor, and even if all individuals experienced the same 
physiological sensitivity to stress, population-level growth would decline exactly as observed 
sole due to the variance effect. This introduces a novel, testable hypothesis that can be 
subjected to future empirical validation using single-cell tracking and targeted statistical 
approaches.

Our finding that individual variation in vital rates tends to suppress population growth aligns 
with earlier observations in bacteria \cite{Steiner2012} and is broadly generalisable beyond 
microbial systems. Similar phenomena have been documented in controlled experiments on plants, 
where controlling for environment and genotype revealed that cohorts with high reproductive 
variance exhibited lower-than-expected population growth \cite{Steiner2021}. Conversely, in 
a distinct line of single-cell research, \cite{Genthon2025} demonstrated that fluctuations 
in single-cell growth rates can enhance population growth if slow-growing cells systematically 
divide at smaller sizes than fast-growing cells. 

Continued research is required to characterise how distinct forms of individual variation 
govern population expansion or decline. A decisive factor appears to be the temporal persistence 
of the variation—namely, whether individuals exhibit distinct trait values persistently throughout 
their lifespan or in a rapidly fluctuating manner \cite{Fox2002}. Critically, the variance-driven 
phenomena analysed in this paper depend on stable, long-lasting differences among individuals 
upon which selection can act. Short-term stochastic fluctuations within an individual's lifespan 
are not expected to yield the same population-level consequences unless they manifest as 
significantly different time-averaged values across separate individuals. Extending the models presented 
here to incorporate stochastic dimensions represents a promising avenue to formalise and rigorously 
differentiate these mechanisms.

\appendix
\section{Alternative stress model}
In the main text, environmental stress was modelled as a reduction in the viability of 
daughter cells at birth. While this assumption is biologically natural, it is not unique: 
stress may also act continuously during the lifetime of a cell by increasing the probability 
of death prior to division. Here, we consider this alternative formulation in order to assess the robustness of 
the results obtained in Section~2. 

\begin{figure}[h!]
\begin{center}
\scalebox{0.45}{\includegraphics{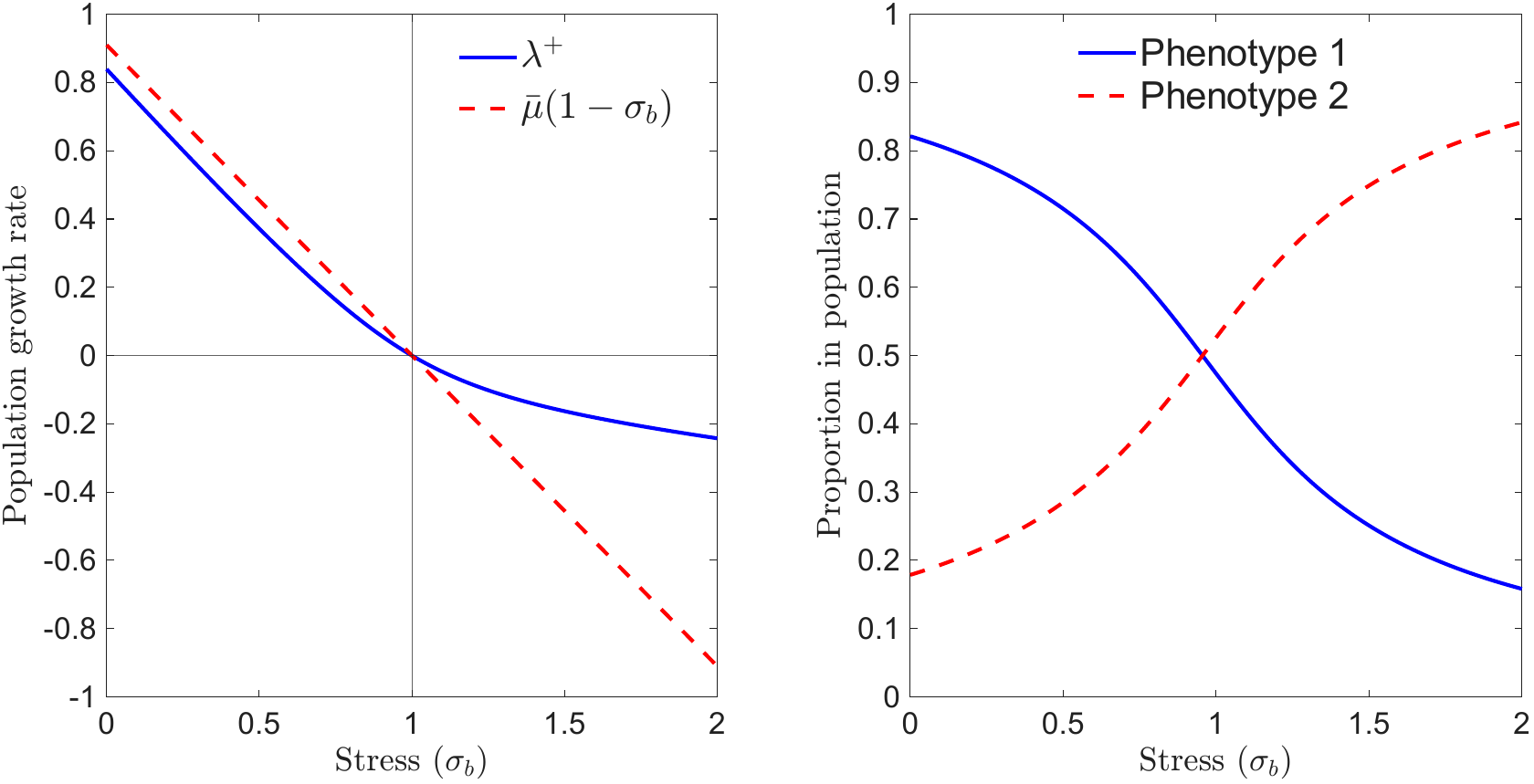}}
\end{center}
\caption{Dependence of population dynamics on stress strength $\sigma_b$ in the 
alternative stress model. (a) Dominant eigenvalue of the system matrix $A_b$ as a 
function of $\sigma_b$, showing the transition from population growth to decay as 
stress increases. The red dotted line indicates the corresponding growth rate 
$\bar{\mu}(1-\sigma_b)$ for the homogeneous (one-phenotype) model with the same 
mean division rate. Comparison with this baseline shows that heterogeneity reduces 
growth for $0 \leq \sigma_b < 1$ and mitigates population decay for $\sigma_b > 1$. 
(b) Corresponding long-time proportion of phenotypes 1 and 2, illustrating the shift 
in population composition with increasing stress, with higher stress favouring the 
slower-dividing phenotype.}
\label{fig:ratio_grow_kill_death_stress}
\end{figure}

As in the two-phenotype model of Section~2, the population 
consists of two phenotypes with number densities $N_1(t)$ and $N_2(t)$ at time $t$, 
and corresponding division rates $\mu_1$ and $\mu_2$, with $\mu_2<\mu_1$. 
Upon division, each daughter is assigned phenotype 1 with probability $p$ 
and phenotype 2 with probability $1-p$, independently of the parent. In contrast to the 
main model, stress now acts during the lifetime of each cell by increasing 
its effective death rate prior to division. Under these assumptions, the population 
dynamics are governed by
\begin{equation}
\frac{dN_{1}}{dt}
=
2p(\mu_{1}N_{1}+\mu_{2}N_{2})
-
\mu_{1}(1+\sigma_{b})N_{1},
\label{eq:n1eqn_app}
\end{equation}
\begin{equation}
\frac{dN_{2}}{dt}
=
2(1-p)(\mu_{1}N_{1}+\mu_{2}N_{2})
-
\mu_{2}(1+\sigma_{b})N_{2},
\label{eq:n2eqn_app}
\end{equation}
where $0\leq \sigma_{b}$ measures the strength of the stress. Here, stress acts 
multiplicatively on the baseline division rate, so that each phenotype 
experiences an effective loss rate $\mu_i (1+\sigma_b)$ prior to division. The parameter 
$\sigma_b \geq 0$ therefore quantifies the relative increase in mortality induced by stress, 
with $\sigma_b = 0$ corresponding to a stress-free environment. This system can be 
written in vector form as
\begin{equation}
\frac{d}{dt}
\begin{bmatrix} N_{1} \\ N_{2} \end{bmatrix}
=
\begin{bmatrix}
(2p-1-\sigma_b)\mu_1 & 2p\mu_2 \\
2(1-p)\mu_1 & (1-2p-\sigma_b)\mu_2
\end{bmatrix}
\begin{bmatrix} N_{1} \\ N_{2} \end{bmatrix},
\label{eq:ode2ph_app}
\end{equation}
or equivalently $\dot{\bfa N}=A_b\bfa N$, where $\bfa N(t)=[N_1(t),N_2(t)]^{T}$. For comparison, 
consider the corresponding homogeneous (one-phenotype) model 
with mean division rate $\bar{\mu} = p\mu_1 + (1-p)\mu_2$. In this case, the 
population growth rate is given explicitly by $\bar{\mu}(1-\sigma_b)$. This 
baseline is included in Figure~\ref{fig:ratio_grow_kill_death_stress}(a), and reveals that the 
two-phenotype model exhibits reduced growth when $0 \leq \sigma_b < 1$, 
vanishing growth at $\sigma_b = 1$, and a reduced rate of decay when 
$\sigma_b > 1$, mirroring the behaviour observed in the main text. As in Section~2, 
the asymptotic behaviour is determined by the dominant eigenvalue of the system matrix. 
Despite the different formulation of stress, the same qualitative behaviour is observed: 
heterogeneity in division rates reduces population growth in favourable environments and reduces the rate of 
population decay under sufficiently strong stress. In Figure~\ref{fig:ratio_grow_kill_death_stress}(b) 
we also observe a systematic shift in population composition with increasing 
stress, with higher stress favouring the slower-dividing phenotype. Taken together, 
the agreement between the baseline, the main model, and 
this alternative formulation suggests that the observed dependence on 
heterogeneity is not tied to a specific stress mechanism, but reflects a 
more general principle of selection under variable growth conditions.

\section*{Acknowledgements}
We are gretaful to Nick Colegrave for providing the \emph{Chlamydomonas reinhardtii} data used in Section \ref{sec:bio}.

\bibliography{refs}
\bibliographystyle{agsm}

\end{document}